\documentclass[draft]{agujournal2019}
\usepackage{url} 
\usepackage{lineno}
\usepackage[inline]{trackchanges} 
\usepackage{soul}
\usepackage{amssymb}
\draftfalse

\journalname{JGR: Space Physics}

\newcommand{\hlst}[1]{\sethlcolor{gray}\hl{\emph{[#1]}}\sethlcolor{yellow}}
\renewcommand{\hlst}[1]{}
\renewcommand{\hl}[1]{#1}

\begin{document}

\title{Statistical Characteristics of the Proton Isotropy Boundary}


\authors{C. Wilkins\affil{1}, V. Angelopoulos\affil{1}, A. Artemyev\affil{1,2}, A. Runov\affil{1}, X-J. Zhang\affil{3}, J. Liu\affil{1}, E. Tsai\affil{1}}

\affiliation{1}{Department of Earth, Planetary, and Space Sciences, University of California, Los Angeles, Los Angeles, California, USA}
\affiliation{2}{Space Research Institute, RAS, Moscow, Russia}
\affiliation{3}{Department of Physics, University of Texas at Dallas, Richardson, Texas, USA}

\correspondingauthor{Colin Wilkins}{colinwilkins@ucla.edu}

\begin{keypoints}
\item We statistically characterize the properties and associated precipitating energy flux of 50 keV to $\sim$2 MeV proton isotropy boundaries (IBs)
\item Proton IBs occur in $>$90\% of auroral zone flybys within 19-03 MLT, typically spanning 64$^\circ$-66$^\circ$ in latitude (L of 5-6), varying with activity
\item IB-associated protons account for 50\%–100\% of nightside high-latitude $\geq$50 keV precipitation, typically 100s of MW, but can exceed 10 GW
\end{keypoints}

\begin{abstract}
Using particle data from the ELFIN satellites, we present a statistical study of 284 proton isotropy boundary events on the nightside magnetosphere, characterizing their occurrence and distribution in local time, latitude (L-shell), energy, and precipitating energy flux, as a function of geomagnetic activity. For a given charged particle species and energy, its isotropy boundary (IB) is the magnetic latitude poleward of which persistently isotropic pitch-angle distributions ($J_{prec}/J_{perp}\sim 1$) are first observed to occur. This isotropization is interpreted as resulting from magnetic field-line curvature (FLC) scattering in the equatorial magnetosphere. We find that proton IBs are observed under all observed activity levels, spanning 16 to 05 MLT with $\sim$100\% occurrence between 19 and 03 MLT, trending toward 60\% at dawn/dusk. These results are also compared with electron IB properties observed using ELFIN, where we find similar trends across local time and activity, with the onset in $\geq$50 keV proton IB occurring on average 2 L-shells lower, and providing between 3 and 10 times as much precipitating power. Proton IBs typically span $64^\circ$-$66^\circ$ in magnetic latitude (5-6 in L-shell), corresponding to the outer edge of the ring current, tending toward lower IGRF latitudes as geomagnetic activity increases. The IBs were found to commonly occur 0.3-2.1 Re beyond the plasmapause. Proton IBs typically span $<$50 keV to $\sim$1 MeV in energy, maximizing near 22 MLT, and decreasing to a typical upper limit of 300-400 keV toward dawn and dusk, with peak observed isotropic energy increasing by $\sim$500 keV during active intervals. These results suggest that FLC in the vicinity of IBs can provide a substantial depletion mechanism for energetic protons, with the total nightside precipitating power from FLC-scattering found to be on the order of 100 MW, at times $\geq$10 GW.
\end{abstract}

\section*{Plain Language Summary}
In Earth's magnetosphere, energetic protons and electrons are often trapped by the geomagnetic field. However, when the equatorial magnetic field becomes too weak or curved to maintain such trapping, these particles can be lost to the atmosphere, giving up their energy as precipitation. We focus on a particular location in the ionosphere at which this occurs, known as the isotropy boundary (IB), for which the number fluxes of precipitating and magnetically reflected (trapped) first become equal. At this latitude, mapped to the geomagnetic equator, the particle's equatorial gyro-radius is of comparable spatial scale to the background magnetic field variations, allowing it to scatter in its pitch angle with respect to the field direction. This randomizes the affected particles in pitch angle, and many thus precipitate. Here, we use observations from polar Low Earth Orbit to characterize statistically the properties of these IBs, such as their occurrence rate versus longitude and latitude, as well as the associated energy they deposit into the atmosphere at different geomagnetic activity levels. This ultimately allows for better understanding and predictive capability of space weather effects, as well as for improvement in magnetic field and electron flux models.

\section{Introduction}
\subsection{Background}
A primary objective in the study of Earth's magnetosphere-ionosphere system is to characterize the sources and mechanisms of energetic charged-particle scattering and precipitation into the atmosphere. One such process is that of magnetic field-line curvature (FLC) scattering (occasionally referred to as ``current-sheet scattering''), which occurs when the gyro-radius of energetic particles approaches length scales comparable to gradients in the background magnetic field direction. This situation commonly arises in the magnetosphere, such as in the magnetotail, where the equatorial field vector often reduces to only a small northward neutral sheet $B_z$ component, accompanied by a sharp radius of magnetic curvature ($R_C$) owing to the transition into straightened earthward and tailward lobe field lines separated abruptly by the cross-tail current sheet. When particles with perpendicular kinetic energies (and therefore gyro-radii) in these regions exceed a characteristic minimum threshold, they can be impulsively scattered in pitch-angle. The result is that energetic particle distributions drifting the curved-field region are stochastically scattered, producing an isotropic distribution of equal number fluxes of precipitating ($J_{prec}$) and locally-mirroring ($J_{perp}$) particles, i.e., $J_{prec}/J_{perp}\sim 1$ \cite{Gray1982,Sergeev1982,Birmingham1984,Delcourt1996,Martin2000}. Under this scenario, upon exiting the scattering region, the particles continue their motion along the instantaneous field line, where they may map into the ionosphere, becoming observable to low-altitude spacecraft. The absolute magnetic latitude poleward of the equator at which such isotropized fluxes are first observed is known as the isotropy boundary (IB) for that particular particle species, energy, and magnetic local time (MLT) \cite{Sergeev1983}.

Isotropy boundaries and associated FLC-scattering are geophysically important for several reasons. For one, they present a robust multi-species pitch-angle isotropization mechanism, affecting electrons, protons, and heavier ions---potentially all simultaneously---with the ability to be active in a large fraction of the magnetosphere in the strong-diffusion limit, under any geomagnetic conditions, dependent only on the large-scale background magnetic field configuration \cite{Sergeev1993}. Because this process scatters particles into the bounce loss-cone, it can lead to the direct precipitation and loss of energetic particles (keVs and up) throughout extended magnetospheric regions, including within the ring current, radiation belts, and plasma sheet \cite{Imhof1979,Sergeev1983,Imhof1997,Yue2014,Selesnick2023,Wilkins2023}. Further, owing to the rapid pitch-angle diffusion on the order of single bounce periods, particles undergoing such scattering provide a near-instantaneous tracer at Low Earth Orbit (LEO) of the distant equatorial magnetic field strength and curvature connected to the particular observed field-lines. By rapidly sampling the IB of different energy and species particles along the satellite orbit track, it is possible to create latitudinal profiles of the instantaneous equatorial magnetic field. This provides a way to determine the magnetotail configuration at any given time, as well as more generally the state of the magnetosphere, and to improve associated modeling \cite{Sergeev1993,Newell1998,Shevchenko2010}. 


Studies of isotropy boundaries have utilized a combination of simulations and in-situ/ground observations to establish their species-, energy-, and geomagnetic activity-dependent behavior. These works have aimed to investigate several key features, including: the occurrence of IBs in magnetic latitude (or L-shell) and local time; the static or time-varying equatorial magnetic field configuration at the IB-mapped scattering location (e.g., $R_C$, $B_z$, and the ``$\kappa$'' parameter described later) and their interpretations as open/closed drift shells and field-line boundaries; and the precipitation/loss of energetic particles into the atmosphere due to FLC scattering in the vicinity of the IB. Simulations have primarily been carried out either by particle tracing through the magnetic equator using analytical field models or global fluid codes, or by estimating the effect of these crossings as a quasi-linear diffusion process \cite{Gray1982,Sergeev1982,Birmingham1984,Delcourt1996,Martin2000,Young2002,Young2008}. For protons, a number of in-situ observations of FLC/IBs have been reported (e.g., \citeA{Imhof1977,Sergeev1983,Sergeev1993,Newell1998,Ganushkina2005,Shevchenko2010,Yue2014,Ilie2015,Sergeev2015a,Sergeev2015b,Dubyagin2018}). Many of these works focused on observations from LEO, considering IBs in the few keVs to low 100s of keV range, relying heavily on the the Medium Energy Proton and Electron Detector (MEPED) instrument on-board the NOAA Polar Operational Environmental Satellites (POES) series of observatories, as well as similar instrumentation on-board the Defense Meteorological Satellite Program (DMSP) observatories, with some studies using ground observations such as CANOPUS to observe precipitation signatures (e.g., \citeA{Donovan2003}). In some cases, conjunctions between LEO and equatorial satellites, such as the Time-History of Events and Macroscale Interactions during Substorms (THEMIS) mission, have been achieved \cite{Shevchenko2010}, although direct observation of FLC/IBs by equatorial spacecraft presents a major technical hurdle for several reasons. For one, the IB is a localized and dynamic structure requiring equatorial spacecraft be within a specific (a priori unknown) 1-2 Re radial distance at the right time, with no way to distinguish whether the instantaneous satellite position corresponds to the actual radial onset location for isotropy, versus a further distance for which FLC remains active (i.e., mapping poleward of the IB in the ionosphere). Further, owing to the typical equatorial local loss-cone size being on the order of a degree, instruments must provide very high pitch-angle resolution. Beyond the issues faced in observing proton IBs, electron IB observations have been even more challenging due to the more distantly-mapped and variable nature of electron FLC scattering, with selected observations have been reported and compared with other scattering processes (e.g., \citeA{Imhof1977, Imhof1979,Shevchenko2010,Sergeev2018, Capannolo2022}). Until recently, the only comprehensive energy-spatial statistics for electron IBs had been compiled by \citeA{Imhof1997} for a limited latitudinal range (57$^\circ$ inclination orbit), followed by more detailed energy-latitude-MLT statistics from \citeA{Wilkins2023} using the polar-orbiting ELFIN satellites.

In this work, we present results in the spirit of \citeA{Sergeev1993} and \citeA{Sergeev2015a} by reporting novel IB observations of energetic protons (H$^+$) between 50 keV and several MeV based on particle data from the ELFIN mission \cite{Angelopoulos2020}. These protons originate from both ionospheric outflows and directly from solar wind flux penetration onto Earth-connected field lines (e.g. by magnetopause reconnection) \cite{Daglis1999}. Through various energization mechanisms, a fraction attain energies a few keVs and higher, appearing throughout the magnetotail during injections \cite{Gabrielse2014,Liu2016}, where the field curvature can be significant. These protons are susceptible to FLC-scattering, owing to their larger gyro-radius approaching the scale of the background field variations. For particles within the near-Earth ring current the plasma sheet, a fraction of particles often attain even higher energies, such as 10s of keVs (plasma sheet), with 100s of keV fluxes appearing during injections, and even MeVs in the ring current. As we demonstrate in the results, the IBs typically occur near the outer edge of the ring current, and produce isotropic fluxes over radial distances penetrating into the inner edge of the plasma sheet. This is important for implications of magnetospheric ion transport and loss during for example during storms and substorms, for which particle fluxes in the plasma sheet and or ring current can be significantly enhanced, and for which FLC scattering may precipitate a large fraction of energy flux into the atmosphere. This can significantly affect the ionosphere, such as by enhancing nightside conductivity via impact ionization along the affected field-lines. In addition, the role of curvature scattering in the evolution of the ring current at both storm- and non-storm times has recently been of significant interest---both as a precipitation source, and as a mechanism to explain observed deviations Dst decay rates versus modeled predictions, with several simulation works having directly- or indirectly-inferred the need for an additional scattering mechanism such as FLC to account for observations \cite{Ebihara2003,Ebihara2019,Yu2020,Dubyagin2020,Chen2021,Eshetu2021,Cao2023}. 

We aim to more fully assess the energy-dependent geophysical properties of proton IBs and associated poleward FLC-scattering by focusing on several observable characteristics: occurrence rates in magnetic local time (MLT); energy- and activity-dependent distribution in magnetic latitude (MLAT) or equivalently L-shell; distribution of energy-integral total and relative precipitating fluxes; and the poleward latitudinal extent of the region presumed to be dominated by FLC-scattering, and its equatorial separation from the plasmapause. Compared to previous studies, our results using ELFIN provide significantly higher energy and latitudinal coverage, revealing the more complete effect and latitudinal extent of FLC-scattered protons, including for moderate strength storms (Dst $\sim$ -50 nT), as well as the fine-scale differential/wide-energy features of proton IBs, which have remained similarly unconstrained outside of simulations or specific case studies. At storm time, observations have previously only been reported for limited energy channels (e.g., 30 keV in \citeA{Newell1998} and \citeA{Dubyagin2018}), or for limited latitudinal coverage (e.g. \citeA{Yue2014}), necessitating further investigation. This was due in part to sensitivity limitations in those detectors at higher energy channels, as well as other orbital and instrumental constraints, which ELFIN can help to overcome.

Additionally, comparison of the statistical properties between electron and proton IBs has remained limited, owing to scant statistics on electron IBs due to the limited observing capabilities of past missions. In this work, building on the recent ELFIN electron IB statistics of \citeA{Wilkins2023}, we provide the first large-sample comparison of their properties over a wide range of energy, MLAT, MLT, and geomagnetic activity levels, as well as relative amounts of precipitating power from FLC-caused precipitation.

\subsection{Model of Energetic Particle Isotropization and Boundary Formation}
Throughout much of the magnetosphere, charged particles travel along magnetic field lines, permitting guiding-center drift and bounce motions with associated adiabatic invariants. However under certain conditions, these invariants can be violated. One such breakdown results in the formation of isotropy boundaries: the abrupt variation of a particle's first adiabatic invariant ($\mu$) upon encountering a region of sufficiently weak/sharply-curved magnetic field geometry. The condition for this breakdown is the particle's equatorial gyro-radius ($r_L$) beginning to approach the scale length of magnetic field variations, quantified by the curvature radius ($R_C$) of the local equatorial field configuration, where $R_C = (\hat{\bf b}\cdot \nabla )\hat{\bf b}$, and $\hat{\bf b} = {\bf B}/B$ is the unit tangent vector to the magnetic field $\bf B$. Such a scenario occurs for example at the center of current sheets, including Earth's cross-tail current, where only a $B_z$ component from the Earth's internal magnetic field exists (commonly referred to as the ``neutral sheet''). 

In the preceding equatorial crossing scenario, the effect of the interaction between the particle and field depends highly on the ratio of the equatorial curvature radius to gyro-radius, defined in prevailing literature as $\kappa^2 = R_C/r_L$. When $\kappa^2 \gg 10$, typical adiabatic guiding-center motion results, with conservation of $\mu$, possibly imparting a change in gyro-phase \cite{Gray1982}. The range of $\kappa^2 \lesssim 1$ results in Speiser-like motion \cite{Speiser1965,Martin2000}, while $3 \lesssim \kappa^2 \lesssim \kappa^2_{cr}$ results in pitch-angle scattering in the strong-diffusion limit \cite{Sergeev1983,Sergeev1993,Delcourt1996,Martin2000}, producing isotropized particle fluxes, where $\kappa^2_{cr}$ is a critical parameter dependent on the geometry within and outside the equatorial region. The quantity $\kappa^2_{cr}$ is typically taken to have an \emph{a priori} value of 8 based on the assumption of a Harris-type current sheet with constant $B_{normal}$ \cite{Gray1982}, but in observations has been reported to span the range of 3 to 33 \cite{Sergeev2015a,Ilie2015}. Additionally, for particles heavier than electrons (e.g. protons), it has been shown that the effect of curvature off of the equator also affects the scattering \cite{Young2002}, with correction terms proportional to $d^2 R_C/ds^2$ and $d^2 B/ds^2$ affecting the minimum gyro-radius for strong diffusive scattering, where $s$ is the path distance along the field-line following the center of the particle trajectory in the vicinity of the equator. 

The relationship $\kappa^2 = \kappa^2_{cr}$ establishes a minimum gyro-radius required for strong diffusive pitch-angle scattering, which can be re-cast in the form of a minimum particle perpendicular kinetic energy $E_{min}$. As shown in \citeA{Wilkins2023}, a relativistic form of this equation (excluding the Young et al. off-equator corrections) is given by:
\begin{equation}
    E_{min} = (\gamma_{min}-1)mc^2 = \left[\left(1 + \left[\frac{q B R_C}{\kappa^2_{cr} m c}\right]^2\right)^{1/2} -1 \right]mc^2
\end{equation}
where $\gamma$ is the particle Lorentz factor, $m$ and $q$ are the mass and charge of the particle, respectively, $B$ is the magnetic field strength, $R_C$ is the curvature radius of the magnetic field, and $c$ is the speed of light. 

Figure 1 shows an example of a spatial profile (GSM noon-midnight xz-cut) of the minimum particle energy $E_{min}$ (Eqn. 1) for particle isotropization due to FLC for electrons (top) and protons (bottom) on 2022-08-04 at 06:08 UT, using the \citeA{Tsyganenko1989} external field model and the International Geomagnetic Reference Field (IGRF) internal field model \cite{Alken2021} to evaluate the magnetic field strength and curvature in space. For both electrons and protons, it can be seen that along any given field line, the lowest energy required for scattering occurs at the geomagnetic equator. The absolute magnetic latitude corresponding to the onset in isotropization by FLC scattering defines the IB. The plot shows that the minimum energy decreases monotonically in tailward distance along the magnetic equator. This monotonic decrease in scattering energy with distance results in the classical feature of IBs: their energy-latitude dispersion as observed in LEO. 

For purposes of IB event identification in this study, we rely on the IB energy-latitude dispersion signatures compatible with the theoretical model of Fig. 1. We note that the accuracy of predictions such as those in Fig. 1 are fundamentally limited by the magnetic field model and choice of $\kappa^2_{cr}$ parameter. This may result in appreciable deviations from the true IB latitude or apparent energy-latitude dispersion; however, it typically does not reverse the monotonicity (i.e., higher energy particles are still isotropized at lower latitudes), and we thus still look for these signatures as identifiers of IBs. We note also that the model in Fig. 1 predicts that at any latitude poleward of an IB, FLC will continue to isotropize particles above that minimum energy, leading to an extended region of isotropic precipitation. As we demonstrate in the the Results section, IBs often exhibit strong particle precipitation in a localized region in their immediate poleward vicinity. To focus on this region separately from plasma sheet fluxes (also isotropic), we impose an operational poleward latitude cutoff at the inner edge of the plasma sheet in calculations of IB-associated precipitation, with the implementation details described in the Methods section.

\subsection{Outline}
Following this introduction section, we next describe the ELFIN mission and its dataset, along with the methodology for identifying and characterizing IB events. We then present statistical results of proton and electron IB characteristics, and a summary and discussion of those results.

\section{Methods and dataset}
\subsection{ELFIN instruments and data}
    To form the dataset used in this study, we make use of electron and proton particle data from the ELFIN mission \cite{Angelopoulos2020}, composed of two small satellites (3U+ CubeSats) in polar Low Earth Orbit (LEO). For the majority of the mission, the satellites were at $\sim$450 km altitude, drifting approximately 1 hour in local time per month, with a slight difference in altitude allowing the two satellites to overlap in local time every 6 months. Over the final 6 months of the mission (March-September 2022), the altitudes decayed rapidly from 420 km to roughly 300 km due to atmospheric drag, followed by terminal reentry.
    
    The primary instruments were a pair of energetic particle detectors for electrons (EPD-E) and protons (EPD-I), nominally measuring differential-directional fluxes of 50 keV to 5 MeV electrons (geometric factor 0.15 cm$^2$-str) and protons (0.01 cm$^2$-str), and a boom-deployed 3-axis Fluxgate Magnetometer (FGM) sampling the background field at 80 samples/sec. The satellites were spinning with a period of 2.8 s in a plane nominally containing the parallel and perpendicular directions to the background magnetic field, allowing for on-board spin phase determination using instantaneous FGM readings. Each spin period was subdivided into 16 consecutive spin sectors, providing $\Delta T \sim 175$ ms time resolution per sector. Proton contamination on the EPD-E was mitigated using an aperture foil, while electron contamination on the EPD-I was mitigated using a magnetic broom filter and two-detector anti-coincidence logic. Contamination by side-penetrating particles was mitigated by a combination of dense shielding surrounding the look directions and knife-edge surfaces to limit the field of view via specular reflection of particles, as well as detector stack coincidence logic.
    
    Particle counts and energies were determined using digitally-sampled pulse-height analysis (PHA) on the output of a pulse-shaping pre-amplifier configured to sense silicon detector impact events. The EPD-I did not possess the capability to resolve ion composition, although we assume all ion counts are protons in the context of this work. While this presents the possibility of false counting, the data were collected during non-strong storm intervals (min Dst $>$ -50 nT) in which proton fluxes are expected to dominate all other energetic ($>$50 keV) species, such as $O^+$ found in the intense storm-time ring current \cite{Daglis1999,Yue2019}. The particle loss cones were computed locally at every timeseries data step by tracing the instantaneous IGRF field at the spacecraft location poleward to a bounce altitude of 100 km, with any such particles with mirror point at or below this altitude considered lost.

    \subsection{Formative events in this study}
    Events in this study are composed of 6-7 minute ELFIN science zone collections, typically spanning $\pm 55^\circ$ to $\pm 80^\circ$ in IGRF magnetic latitude in a single hemisphere, traversing field lines which map into the outer radiation belt, ring current, plasma sheet, and polar cap. For an ELFIN collection to qualify as a candidate event for IB determination, we required that the observations covered a range of pitch-angles sufficient to resolve precipitating to mirroring flux ratios, and that the latitudinal coverage of the collection spanned at least IGRF $3<L<6$ for protons and IGRF $3<L<8$ for electrons, to ensure that higher-latitude IBs were not artificially under-represented. The data were also manually inspected for any potential instrumental issues, which such cases being removed. We refer to ELFIN collections which meet these validation criteria as Qualifying Science Zones (QSZs), which resulted in a reduction of 6500 electron collections to 4150 equivalent QSZs, and reduced 337 proton collections to 300 equivalent QSZs.
    
    A prototypical science zone collection on 2022-08-19 containing both electron and proton IBs observed by ELFIN-B is presented in Figure 2. The satellite is observed to begin in the southern hemisphere around -55$^\circ$ in magnetic latitude around 23.5 hours MLT, collecting for $\sim$7 minutes, spanning latitudes including the outer radiation belt, ring current and plasma sheet. Panels (a) and (b) show the energy flux spectrograms of locally-mirroring (perpendicular) and precipitating (parallel) electrons respectively, with Panel (c) showing the precipitating-to-trapped flux ratio $R_I = J_{prec}/J_{perp}$ for electrons. Between 0605 and 0608 UT, the spacecraft sees well-populated locally-mirroring energetic electron fluxes with a low isotropy ratio, corresponding to dynamics controlled by the outer radiation belt. However, around 0608 UT a rapid rise in isotropy from $R_I < 0.6$ to $R_I \sim 1$ is observed across all energy channels, with accompanying energy-latitude dispersion qualitatively consistent with the approximately monotonic dispersion signature of isotropy boundary formation by FLC scattering (as in Fig. 1). Panel (d) shows the model predicted IB locations using T89 (blue curve) versus the observed IBs (red curve), where it is observed that T89 predicted a similar energy-latitude dispersion slope, but with a slight overestimation of the onset latitude mapped to the tail. (Deviations between field model predictions and IB observations present a potentially rich line of investigations, and are beyond the scope of this study.) Panel (e) shows the instantaneous integral (over energy and pitch-angle) precipitating energy-flux of isotropic $\geq$50 keV electrons at the spacecraft location, where it can be seen that the region within and immediately poleward of the IB crossing produces significant power on the order of 0.1 to 1.0 erg/cm$^2$-s, and is the dominant region of electron precipitation in this specific event, contributing 71\% of the total high-latitude energetic electron precipitation observable by ELFIN.

    Panels (f) through (k) are corresponding stack plot for protons equivalent to Panels (a) through (e) for electrons over the same time interval and latitude range. In this case, at lower latitude, there are nearly zero counts of either precipitating or mirroring protons, corresponding to a typical off-equator pitch-angle distribution in the ring current. However similar to electrons, near 06:06:30 UT, an abrupt rise in anisotropic to isotropic fluxes are observed across all proton energies, akin to the energy-latitude dispersion predicted for FLC-scattered protons in Fig. 1. In this case, the model prediction is very close to observation (although not all instances have such good agreement --- a topic similarly-ripe for future study). As with the electron IB, precipitating energy fluxes from isotropic ($>$50 keV) protons rise significantly relative to other latitudes, contributing 90\% of the total high-latitude energetic proton precipitation, although distributed over a wide range of latitude. Lastly, Panel (l) shows the local IGRF field at the spacecraft location for reference.
    
    \subsection{Event classification}
    ELFIN electron and proton collections such as shown in Fig. 2 form the basis for event characterization in this study. To form the event statistics, each qualifying science zone in the database was manually inspected for the presence of an electron (IBe) and or proton (IBp) isotropy boundary. The IB presence was assessed by subjecting the isotropy ratio $R_I = J_{prec}/J_{perp}$ to the following criteria: a poleward transition from anisotropic to isotropic fluxes (defined as passing from $R_I < 0.6$ to $R_I > 0.6$), followed by $R_I \sim 1$ for at least 50\% of the poleward spin periods; 2) at least three energy channels present in the onset of isotropy; and 3) energy-latitude dispersion across all energy channels consistent with FLC (e.g. as in Fig. 1). In the context of IB identification, the quantity $R_I$ was not evaluated for low-count rate conditions (e.g. for 0 counts, or for $dQ/Q \geq 1$, where dQ is the counting uncertainty in flux Q).
    
    We note that while condition 3) was typically satisfied for electron IBs, the g-factor of the EPD-I instrument did not always permit sensitivity of the highest proton energy channels in an observed IB, which can sometimes lead to energy-latitude dispersion opposite to expectation for the highest energies in the crossing. To account for this behavior, events in which higher energy channels were isotropized at higher latitudes (i.e. opposite expected dispersion) were still counted as containing an IBp, provided: a) there is a continuous set of corresponding lower energy channels in the onset of isotropy which exhibit the expected dispersion; and b) the higher energy fluxes transition abruptly from below the 1-count level to populated isotropy fluxes (i.e., fluxes at these energies may actually have been isotropic at lower latitudes than observed, but the count rates were too low to resolve).

    Given the multi-year collection of electron data, all MLTs were sampled with a minimum of 50 events per hourly MLT bin. For protons, data was only available on the nightside in a normal-like distribution between 1700 and 0400 hours (peaked between 22-23 MLT), with each proton collection accompanied by a simultaneous electron data for comparison. This resulted in roughly an order of magnitude more electron than proton QSZs for the statistics. QSZs which met the specified event criteria were marked as containing an IB for each respective species, and their latitude of onset for each energy channel were recorded alongside the net IB/FLC-associated precipitation each species deposited into the atmosphere (relative and absolute precipitating power) integrated over latitude, energy, and the 100 km bounce loss cone, under the assumption of negligible back-scatter. An operational poleward latitudinal cutoff was imposed on the precipitation calculations to limit the possibility of the scattering having been caused by processes other than FLC, since ELFIN could not directly sense conditions at the distant equatorial foot-points (e.g. presence of waves). This cutoff was selected based on the notion that the plasma sheet inner edge is always poleward of the IB, and that processes within this region may or may not be dominated by FLC. Based on the observations and methods in \citeA{Christon1989,Christon1991}, we make use of the fact that the plasma sheet typically has a high-energy ceiling that can be used to assess whether a particular field line maps into the plasma sheet relative to those along the same orbit track. We found this edge to most typically correspond to the cut-off in 300 keV omni-directional flux in ELFIN data, which we evaluate for every event to determine the poleward latitudinal cut-off for FLC-based precipitation calculations. (See the Results section for justification of the 300 keV reference value.)

    Additionally, for each event, the equatorial separation between proton IBs and the plasmapause was also computed. This was done to account for potential interactivity between energetic isotropic protons and the plasmasphere, which can result in instabilities and wave generation, as well as provide a measure of the state of magnetospheric convection and compression. For each IBp event, the L-shell difference between the IB and plasmapause was computed using the \citeA{OBrien2003} model using the ``36 hour AE maximum'' method, taking the most equatorward (typically highest-energy) portion of the IB as the reference for computing separation in IGRF L-shell.

\section{Results}
\subsection{Occurrence and spatial distribution}
    The statistical results of this study were obtained through manual inspection of every ELFIN qualifying science zone using the method described in the preceding section for both electrons and protons. The statistics of electron IBs have been updated to include more than double the events considered in Wilkins et al. (2023), and their average properties are over-plotted with proton results for comparison. Of the 300 proton QSZs, 284 were found to contain proton IBs. Figure 3 shows the scatter plot in the GSM xy-plane of the equatorial field-line trace projections of electron (top) and proton IB (bottom) locations corresponding to the minimum (typically highest-energy) and maximum (typically lowest-energy) portions of the dispersion region. It can be seen that both electron and proton IBs are widely distributed across the nightside, with protons generally occurring at much lower projected L-shells. We note that the true MLT range of proton IBs extends beyond what is shown in the scatter plot; the lack of reported occurrence rates at these MLTs is simply due to lack of satellite coverage there.
    
    Figure 4 (top) shows the distribution in MLT ($\Delta MLT = 1$ hr) of proton QSZs (blue bars) and IBp events (orange bars), with the occurrence rates per MLT bin (ratio of orange to blue bars) plotted in red. The plot shows that proton IBs are observed by ELFIN between 90\% to $\sim$100\% of the time between 1900 and 0300 MLT, with occurrence rates trending downward toward the flanks (ELFIN again had too few collections to determine the proton occurrence rates there or on the dayside). To reduce mapping uncertainties, we applied a $\pm$1 hour averaging filter to the occurrence rate curve, and eliminated bins with $<$5 events. The equivalent occurrence rate for electron IBs are shown in purple, and are found by ELFIN to jointly occur up to 80\% of the time with proton IBs across much of the nightside.
    
    Fig. 4 middle and bottom show the MLT-aggregated distribution in IGRF L-shell and magnetic latitude (MLAT) respectively of the proton IB minimum and maximum projected boundaries of the dispersion region. The most common bounds are between IGRF L=5 to L=6, or 64$^\circ$ to 66$^\circ$ in MLAT, which covers a range typically associated with the higher-latitude portion of the ring current. This is in contrast to electron IBs, which were found in \citeA{Wilkins2023} to be between L=7 to L=8, or $\Delta L = 2$ higher in typical onset separation.
    
    Figure 5 shows the activity dependence of several IBp properties, where ``active'' intervals are defined as 1 hour average AE above 200 nT preceding the ELFIN collection, which split the dataset roughly in half. Panel (a) shows the distribution of minimum and maximum proton energy associated with the IB dispersion region, typically representing a continuously-populated isotropic particle spectrum between those energies. Across all activity and represented MLTs, the minimum isotropic energy is the lowest energy ELFIN can resolve (50 keV), suggesting the IBs likely extend to lower energies. However the maximum isotropic energy exhibits a clear dependence with both MLT and activity. At non-active time, the typical peak proton energy is around 600 keV between 23 to 01 MLT, dropping to 300-400 keV near dusk/dawn. At active times, the mean peak energy reaches 900 keV to 1 MeV, shifting in MLT to 21-23 hours, dropping to 600-700 keV between 01-02 MLT. 
     
    Panels (b) and (c) show the proton IB dispersion region min/max L-shell and MLAT distributions. At non-active time, a bowl-shape distribution similar to electron IBs is seen centered closest on 00-01 MLT (anti-symmetric to the 22-23 MLT from electrons). At active times the distribution shifts to pre-midnight and moves radially inward by 1-2 L-shells, peaking at 20-21 MLT, although the low sample size of events in the vicinity makes it impossible to determine whether the trend continues, and whether it is bowl-shaped. Electron IBs as in \citeA{Wilkins2023} exhibit a similar distribution around 20-22 MLT, suggesting they are controlled by the same or similar processes. Panels (d) and (e) show the slope of the proton IB energy dispersion in L-shell and MLAT, vs MLT and activity. Interestingly the proton IBs seem to have much more uniform dispersion than electrons of around 1 L/MeV. At active times the dispersion seems to become sharper across all local times by more than double, reaching 0.4 L/MeV.

    Figure 6 shows the connection between proton IBs and the plasmapause location as determined using the \citeA{OBrien2003} model based. We used to the recent 36 hour AE maximum as the parameter encoding magnetospheric state, separated by quiet and non-quiet intervals as determined by 1 hour average AE above or below 200 nT. The top Panel in Fig. 6 shows the mean and standard deviation of the lowest IGRF L-shell (lowest latitude) portion of the proton IBs, typically corresponding to 500 keV to 1 MeV energies, alongside the equivalent mean and spread in plasmapause L-shell. It can be seen that the quiet-time IBs form a U-shape distribution centered on roughly on 01 UT, while the plasmapause exhibits a somewhat inverted U-shape centered on 21 UT, with a typical mean separation between 1 and 2 Re. At non-quiet times, both the proton IB and plasmapause move equatorward, breaking the apparent U-shape symmetry, and both trending toward lower latitudes at pre-midnight. In this case, the typical separation between the two decreases to between 0.5 and 1 Re. The bottom Panel in Fig. 6 shows the event-wise activity-dependent difference between the instantaneous IB location and plasmapause. It can be seen to confirm the general statistical trend, revealing a symmetric U-shape difference at quiet time centered on 23 UT, while at active time, the U-symmetry disappears, resulting in much closer separation, minimizing in the pre-midnight sector. 

    \subsection{Precipitation associated with proton IBs and poleward FLC}
    As alluded to in the methods section, in order to discuss the precipitation associated with proton IBs and the dynamical transition region dominated by in their immediate poleward vicinity, we found it necessary to introduce an operational definition for the equatorial source region (latitudinal range)  IB-associated curvature scattering. To do this, we used the same approach as in \citeA{Wilkins2023} for the electron ``PS2ORB'' region, which was to regard the inner edge of the plasma sheet (for each species separately) as a conservative proxy for the poleward edge of IB-related curvature scattering. This approach takes advantage of the fact that the electron and proton plasma sheets proper have characteristic upper energy cutoffs, with the specific value depending on instrument sensitivity \cite{Christon1989,Christon1991}. By finding the characteristic energy cutoff for the plasma sheet in ELFIN data, an operational poleward latitudinal cutoff for the FLC scattering region can be found in every individual collection.
        
    Fig. 7 top and middle explore this connection. Top Panel shows the energy-dependent mean and median locations for proton IBs and the location of poleward omni-directional flux dropouts, where it can be seen that there is a near uniform 2-3$^\circ$ difference in latitude between these curves above 200-300 keV, while the difference balloons at lower energies to over 6$^\circ$, suggesting a separate plasma sheet population is encountered. The middle Panel shows the mean and median event-wise difference between the IB and the omni-directional flux dropout vs energy. This plot reveals a similar ballooning trend at energies below 200-300 keV, which, along with the top Panel, suggests 300 keV is a reasonable proxy through which to determine a conservative poleward cutoff of IB precipitation.
        
    Based on the above 300 keV proton flux dropout criterion, Fig. 7 bottom Panel shows the typical width in latitude of the equatorial proton scattering region dominated by FLC vs MLT and activity. At non-active time, the latitudinal extent of the PS2RC interface is typically 2$^\circ$-3$^\circ$ across all available local times. At active times, the PS2RC interface expands across the entire nightside, with a very clear MLT dependence emerging in which the extent grows from 3.5$^\circ$ at 01-02 hours to 6$^\circ$ between 20-21 hours. Because the range of latitudes involved in FLC-dominated scattering typically span the higher-energy portion of the ring current to the inner edge of the proton sheet (by definition), we refer to this region as the ``plasma sheet to ring current`` (or PS2RC) interface, akin to that of the PS2ORB defined for electrons.

    Using the definitions of PS2RC and PS2ORB, we separately characterize the precipitation from FLC acting on protons and electrons on the nightside. We compute the amounts of precipitating power from curvature scattering of each species assuming integral over all loss-cone pitch-angles and observed energies, assuming $2\pi$ steradian incidence on the atmosphere. The fluxes are averaged in time over the range of latitudes in which they were collected, resulting in an average integral energy flux in units of erg/cm$^2$-s for each individual science zone (herein referred to as ``IEflux''), whose total power deposition can be estimated by scaling over the atmospheric area subtended by the orbit track and the particular MLT sector. In order to assess the relative significance of FLC-caused precipitation, we separate the latitudinal range of the ELFIN collections into 1) the entire science zone (nominally 55-80 deg, including outer radiation belt, ring current, and portions of the plasma sheet/polar cap); 2) all latitudes poleward of the IB (up to the same high-latitude boundary as in 1); 3) latitudes solely within the PS2RC (protons)/PS2ORB (electrons). Region 3 represents the amount of precipitation that is presumed to be associated with the IB, and whose ratio can be taken to the total science zone to determine the relative amount of precipitating $>$50 keV power FLC provides compared to the entire high-latitude region (auroral/sub-auroral zone).
        
    Fig. 8 shows the distribution of latitude-averaged integral precipitating proton energy flux in each of the above regions, poleward latitudinal extent ($\Delta \bar{\theta})$ of each of the three regions described above. We find that of these regions, the PS2RC tends to have the highest average integral precipitating $\geq 50$ keV proton energy flux (on the order of $10^{-1}$ to $10^{-2}$ erg/cm$^2$-s), and is typically concentrated to $3.4^\circ$ in latitude, suggesting it is a significant source of precipitation acting over a range of auroral latitudes.
        
    We also investigated the distribution in MLT of energetic proton precipitation due to FLC in the PS2RC, computing the IEflux in that region, as well as the ratio the precipitation ratio to the total high-latitude region. Fig. 9 shows these results, where the top Panel depicts the precipitation ratio compared with the total amount contributed by its own species over the science zone, and the bottom Panel depicts the precipitating IEflux versus local time over all events containing IBs. The mean values in the distribution for protons and electrons are over-plotted as pink and purple lines. In the top Panel, we see that protons exhibit a highly-symmetric distribution in the amount of high-latitude particle power FLC precipitates, peaking between at 57\% 21-23 MLT, and dropping to 20\% approaching dusk/dawn at 19 and 3 MLT, though at times the total power can approach 100\% for any observed MLT. The trend is similar for electrons, except that protons typically provide almost 3 times as much relative power across the nightside. In the bottom Panel, we see a similar symmetric distribution for of IEflux, with protons typically precipitating with an energy flux on the order of $10^{-2}$ erg/cm$^2$-s versus MLT, but at times exceeding 1 erg/cm$^2$-s between 20-01 MLT. Electrons exhibit a similarly symmetric distribution in MLT, with mean IEflux roughly 1/3 of the proton levels.

    Using the precipitating IEflux from Fig. 9 bottom and the PS2RC latitudinal extent from Fig. 7 bottom, we also compute an average total precipitating power across the nightside, along with a lower bound for the maximum power the process may instantaneously provide. Assuming 111 km/$^\circ$ of latitude and $2\pi$ steradian of precipitating solid angle at the atmosphere, we find that the typical precipitating power of isotropic protons by FLC is on the order of 100 MW, but at the most active times may exceed 10 GW across the nightside. This typically exceeds the equivalent power provided by energetic electron FLC precipitation, which in \citeA{Wilkins2023} were found to supply an average 10 MW across the nightside, and exceeding 1 GW at the most active times. 
        
    Lastly, we characterize the MLT-aggregated precipitation versus geomagnetic activity (AE, Dst, Kp). Fig. 10 shows the resulting precipitation ratio distributions (left column) and IEfluxes (right column), versus AE (100 nT bins; 3 hour average), Dst (10 nT bins), and Kp, using the same definitions for the precipitation ratio and net total average precipitating energy flux as in Fig. 8. The mean values of the distribution for protons and electrons are similarly over-plotted in pink and purple, respectively.

    For AE (top Panel), a clear increase in both the precipitation ratio and integral energy flux is observed as AE increases from 0-200 nT (quiet time), with increases above 200 nT (active times), for which proton almost always exceeds 70\% of the high-latitude total, providing between $10^{-2}$ and $10^0$ erg/cm$^s$-s. Electrons follow a similar trend, but are lower in precipitation ratio and IEflux intensity. For Dst (middle), proton precipitation is observed to rise dramatically as Dst decreases from 0 nT, consistent with FLC being an important mechanism affecting the population storm-time ring current-associated energetic protons. Below -30 nT, the precipitation tends to represent between 70\% to 100\% of the total $\geq$50 keV precipitation with average fluxes between 0.1 and 1 erg/cm$^s$-s, dropping to 40\% for Dst$\sim 0$ nT with average flux between $10^{-2.5}$ to $10^{-2}$ erg/cm$^s$-s. As with AE, electrons show a similar trend versus Dst, although they apparently exhibit a slight increase for positive Dst---possibly exhibiting species-dependent behavior during for example Storm Sudden Commencements (SSCs). For Kp (bottom Panel), the trend is similar to AE, with Kp=3 roughly corresponding to AE=200 nT.
    
\section{Summary and discussion}
In this work, we used ELFIN observations from LEO to determine the observational characteristics of 300 proton isotropy boundary events, as well as the associated poleward precipitation from field-line curvature scattering. We found that proton IBs can be well-recognized across the nightside, with $>$90\% occurrence between 19 and 03 UT, typically spanning 64$^\circ$-66$^\circ$ (L of 5-6) with a continuous spectrum of energies from 50 keV up to several MeV, depending on geomagnetic activity. 

The data revealed that proton precipitation associated with IBs accounts for between 50\% to 100\% of the total $>$50 keV proton energy flux deposited into the atmosphere between 55$^\circ$ and 80$^\circ$ latitude in the MLTs surrounding midnight, with typical IEflux between 10$^{-2}$ to 10$^0$ erg/cm$^2$-s, typically extending between 2$^\circ$ and 6$^\circ$ degrees of latitude, depending on MLT and activity, resulting at times in over 10 GW of precipitating power across the nightside. This agrees with and extends empirical observations at lower energies by \citeA{Newell1998}, who found that for 30 keV protons, the peak in auroral proton at these energies precipitation typically occurs in the immediate poleward vicinity of proton IBs, and is therefore potentially a very geophysically important precipitation process. We suggest that the role this precipitation plays in forming aurora and ionospheric ionization enhancements presents a ripe avenue for investigation.

Along these lines, we have also identified that at the inner edge of the plasma sheet, in the dipole to tail transition region where the PS2RC is located, is a region of high importance for tail heating and energy conversion during storms and substorms. We find that its dynamics can be well-captured from low-Earth orbit since it is most often clearly seen in $J_{prec}/J_{perp}$ ratios. The equatorial energy flux can also be tracked from LEO, as the average field-aligned flux within the PS2RC is a good proxy of equatorial flux, since it is well-isotropized by FLC scattering. Therefore our results are a good approximation of the evolution of this region in MLAT. The westward and equatorward progression with activity can be thought of as due to the plasma sheet ion heating, resulting in higher ion pressure moving closer to Earth likely due to plasma sheet injections. This increases the thermal plasma pressure in the near-Earth regions which acts to reduce the equatorial Bz and hence thin the current sheet \cite{Lu2018}, corresponding to equatorial $B_z$ reduction at dusk, as a means of causing the current sheet to become unstable to reconnection at pre-midnight. This likely causes the thin current sheet near the Earth to move even closer to Earth at pre-midnight, as observed. The PS2RC is also important for precipitation and its location, well below the auroral zone, suggests that the ion contribution to net ionization could be comparable to that of electrons, hence they may contribute significantly to ionospheric conductivity. Therefore they may need to be included in conductivity models; an effect that is currently missing from such models, such as those based on the Robinson formula, which only incorporates $\leq$30 keV electrons \cite{Liemohn2020}.

In the context of the ring current and storms, we observed that both the proton IB location and its absolute and relative precipitating power all scale directly with Dst. This is in agreement with \citeA{Dubyagin2018} who showed the variation in 30 keV proton IBs with Dst and dynamic pressure, which we extend to higher energies (up to MeVs). This also supports the notion that FLC scattering of ring current protons presents a very significant means of depleting such fluxes at higher L-shells, potentially affecting their contribution to Dst during storm main phase and recovery. Unfortunately ELFIN was only active during moderate strength events (min. Dst$\sim$-50 nT), which we argue necessitates further investigation using similar observations for more intense storms (min. Dst$<$-100 nT) to determine whether this correlation continues, and the effect that curvature scattering has on hot heavier ions, such as $O^+$, in the ring current.

We also compared proton IB properties with electron IBs observed with ELFIN, finding that at quiet times, proton IBs have peak occurrence in the post-midnight, while electrons appear more frequently in the pre-midnight, separated typically by 1-2 IGRF L-shells for the same energy channel. At active times, IBs of both species tend to shift to lower latitudes, with both tending toward more extreme properties in the pre-midnight region, suggesting that they share common driving, such as the intersection of tail injections with pre-midnight current-sheet thinning. However, proton IBs are found to have their poleward FLC precipitation extend over a larger range of latitudes (3$^\circ$-4$^\circ$), while those of electrons tend to be more concentrated (1$^\circ$-2$^\circ$). Interestingly, proton IBs exhibit extreme growth of their latitudinal scattering region (e.g. as in Fig. 7 bottom) in the pre-midnight during active times, which may be related to the evolution and decay of the partial ring current. IBs of both species have the potential to precipitate significant power; however, protons typically provide $\sim$3 times as much on average, extended more broadly in latitude, while electrons tend to have very significant precipitation in the immediate poleward vicinity of the IB. 

We additionally determined the separation between proton IBs and the plasmapause, which provides insight into their interaction. We found that at quiet time, the two are typically separated by 1-2 Re, reducing to 0.5-1 Re at active time. We note that in 10\% of cases, almost entirely at quiet time, the IB actually occurred at a lower L-shell than the model plasmapause; however, all of these instances are within the error margin of the model, and are likely in very close proximity, but with the proton IB being poleward in reality. This model also does not consider localized features, such as the plasmaspheric plume, and are therefore not captured in the statistics.

While this study aimed to provide a picture of the wider-energy dependent features of proton IBs, it was fundamentally limited in the number of events that could be collected before the spacecraft de-orbited. For this reason, the occurrence rate statistics outside of 19-03 UT may not be trustworthy (notably the rapid occurrence rate decay approaching dawn and dusk). Whether this is a real feature requires further investigation. The observations were also generally limited in that the geometric factor of the ion detector reduced sensitivity to the $\sim$2 MeV level despite the silicon enabling detection of $>$5 MeV protons. For some events (particularly at quiet times, or when $>$500 keV fluxes have been otherwise depleted), this limited our determination of the highest-energy portions of the proton IBs, which likely extend above 2 MeV at fluxes below ELFIN's sensitivity, as well as estimates of the corresponding multi-MeV portion of the energy-latitude dispersion. We recommend further observations along these lines with increased sensitivity payloads to cover a more wide range of energy and MLT, including at the flanks and on the dayside, to more fully constrain their characteristics. Further ground observations to determine the effect of the harder-energy spectrum (MeV and up) precipitation would also be beneficial, as many studies have focused on only the lower (10s of keV) portion of the precipitation from the plasma sheet. 

Lastly, combined with the electron IB observations, we mention that these databases provide a suitable platform from which to derive predictive model of IB properties of both species, given predictors such as magnetospheric state variables (e.g. Dst), instantaneous solar wind coupling (e.g. by IMF $B_z$ and dynamic pressure), and geometric orientation effects such as MLT. Such a model would not only be useful for the prediction of precipitation and space weather, but also in constraining magnetic field models by tuning their average properties to match observations.

\section{Open Research}
The ELFIN data and software used in this study are part of the SPEDAS framework \cite{Angelopoulos2019}, which is freely available to the public at the following url: \url{http://spedas.org/wiki/index.php?title=Main_Page} 






\acknowledgments
This work has been supported by NASA awards NNX14AN68G, 80NSSC19K1439, and NSF grants AGS-1242918 and AGS-2019914. We are grateful to NASA's CubeSat Launch Initiative for ELFIN's successful launch in the desired orbits. We acknowledge early support of ELFIN project by the AFOSR, under its University Nanosat Program, UNP-8 project, contract FA9453-12-D-0285, and by the California Space Grant program. We acknowledge critical contributions of numerous ELFIN undergraduate student interns and volunteers.

\bibliography{bib_ib_paper}

\section*{[Figures]}
\begin{figure}
\noindent\includegraphics[width=\textwidth]{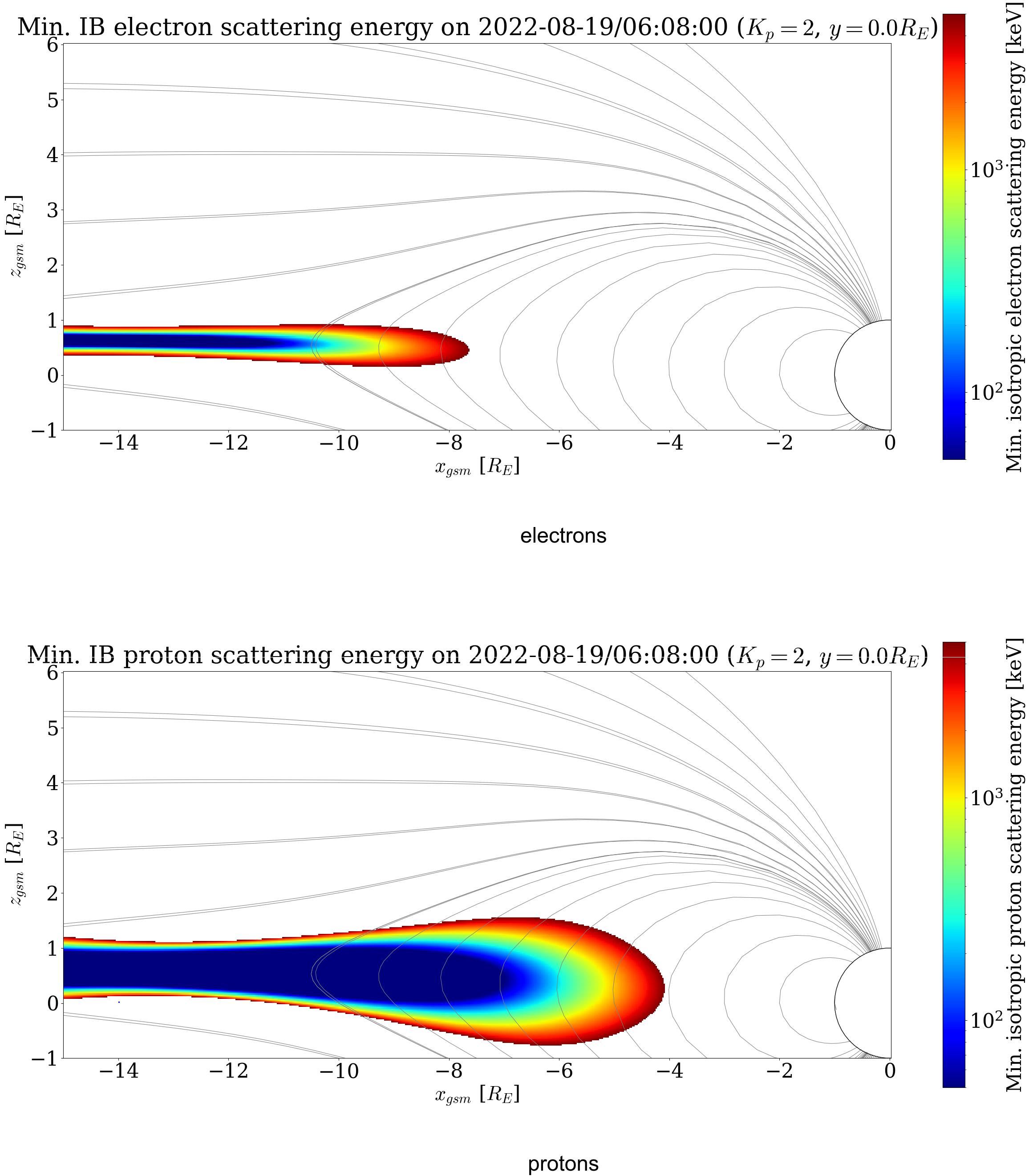}
\caption{Modeled spatial profile (GSM noon-midnight cut) of the minimum kinetic energy required to experience strong-diffusion pitch-angle scattering by field-line curvature (T89 field model), resulting in isotropic distributions observed at LEO (defined in Eqn. 1). The top Panel shows the profile for electrons, while the bottom Panel is for protons. For any given field-line, the IB is nominally defined by the absolute minimum energy along the field line, occurring at the magnetic equator. Electrons experience FLC scattering in a narrow region confined to the equator at L-shells in the typical vicinity of the outer radiation belts and plasma sheet, while protons of the same energy are initially isotropized lower in the vicinity of the ring current. Protons are also observed to experience scattering for a longer arc length of the field line compared to electrons, consistent with the description of \citeA{Young2002}.}
\label{Figure 1.}
\end{figure}

\begin{figure}
\noindent\includegraphics[width=\textwidth]{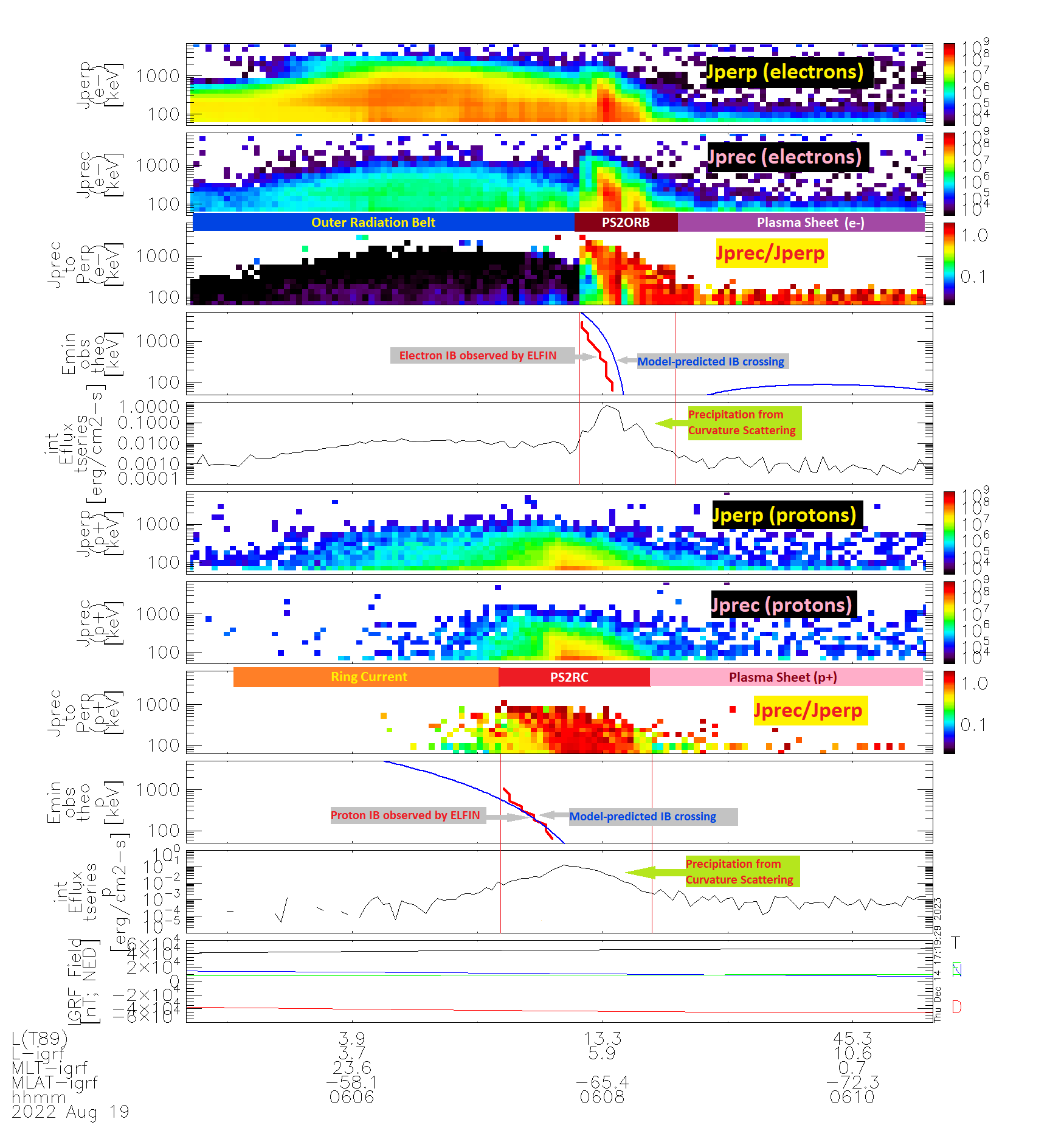}
\caption{Prototypical combined electron+proton IB event observed by ELFIN on 2022-08-19/06:05 UT. Panels (a) and (b) contain the precipitating ($J_{prec}$) and locally-mirroring ($J_{perp}$) energy-time spectrograms of 50 keV to 5 MeV electrons, with Panel (c) being their isotropy ratio $R_I = J_{prec}/J_{perp}$. The spacecraft starts in the southern outer radiation belt around $-55^\circ$ MLAT, and is observed to encounter a rapid rise from low isotropy to $R_I\sim 1$ across all energy channels from 06:07:50 UT to 06:08:10 UT, marking the crossing of 50 keV to 3 MeV electron isotropy boundaries. Panel (d) shows the theoretical model prediction using Eqn. 1 and T89 (blue) compared with the actual observations (red), while Panel (e) shows the instantaneous integral precipitating energy flux, which at times approaches 1 erg/cm$^2$-s. The red bars denote the region considered to be dominated by curvature scattering for electrons (the ``PS2ORB'') and protons (the ``PS2RC''), the transition region between the IB and plasma sheet (see text for definitions). Panels (f)-(k) represent the equivalent quantities for protons as Panel (a)-(e) did for electrons, where it can be seen that the spacecraft begins in the ring current, with highly anisotropic fluxes. A 50 keV to 1 MeV proton IB is then encountered around 06:07:30 at the higher L-shell portion of the ring current ($L\sim5.5$). As expected, it occurs at lower latitudes than the electron IB, and proton FLC scattering region exhibits significant precipitation in the immediate poleward vicinity of the IB, consistent with reports by \citeA{Newell1998} at lower energies.}
\label{Figure 2.}
\end{figure}

\begin{figure}
\noindent\includegraphics[width=\textwidth]{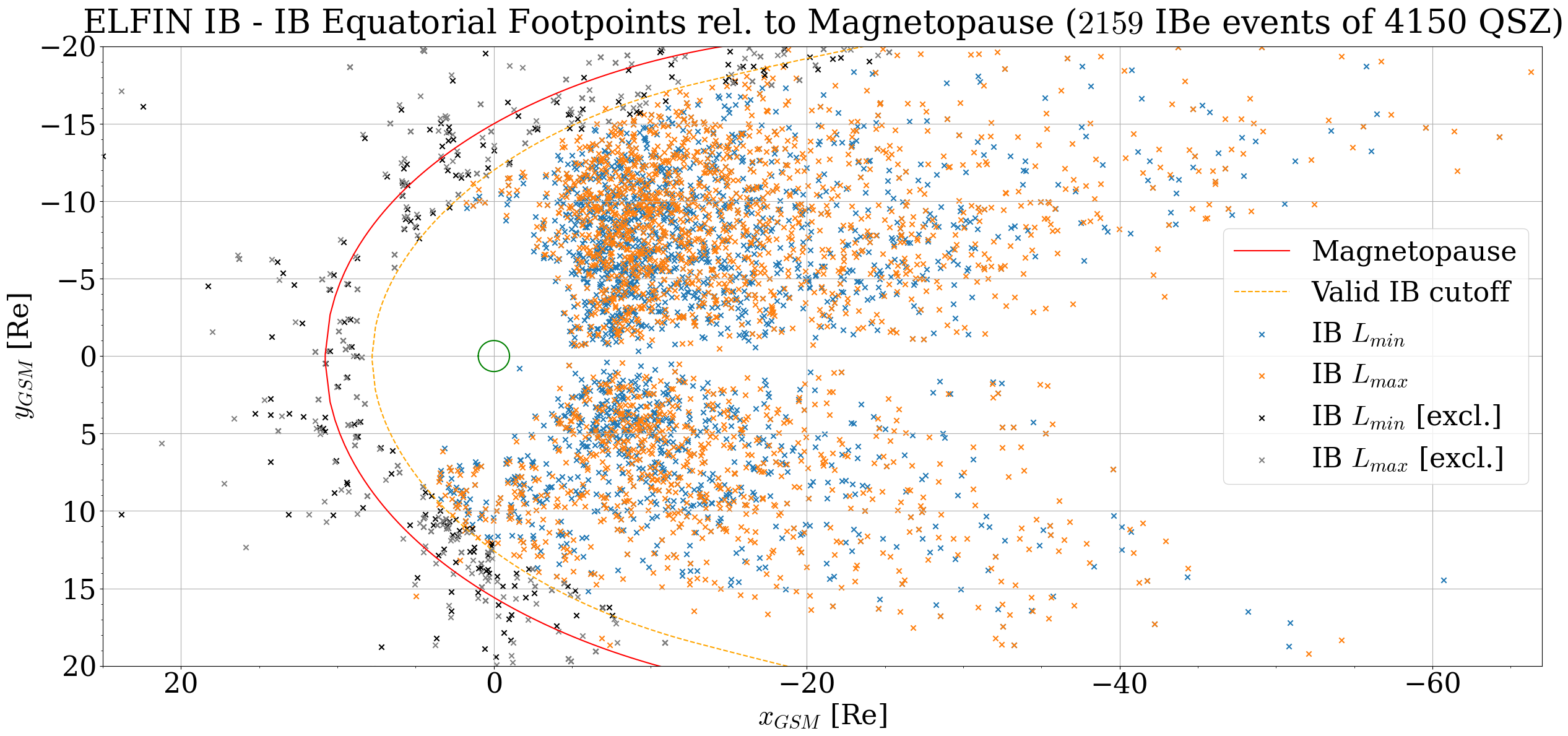}\\
\noindent\includegraphics[width=\textwidth]{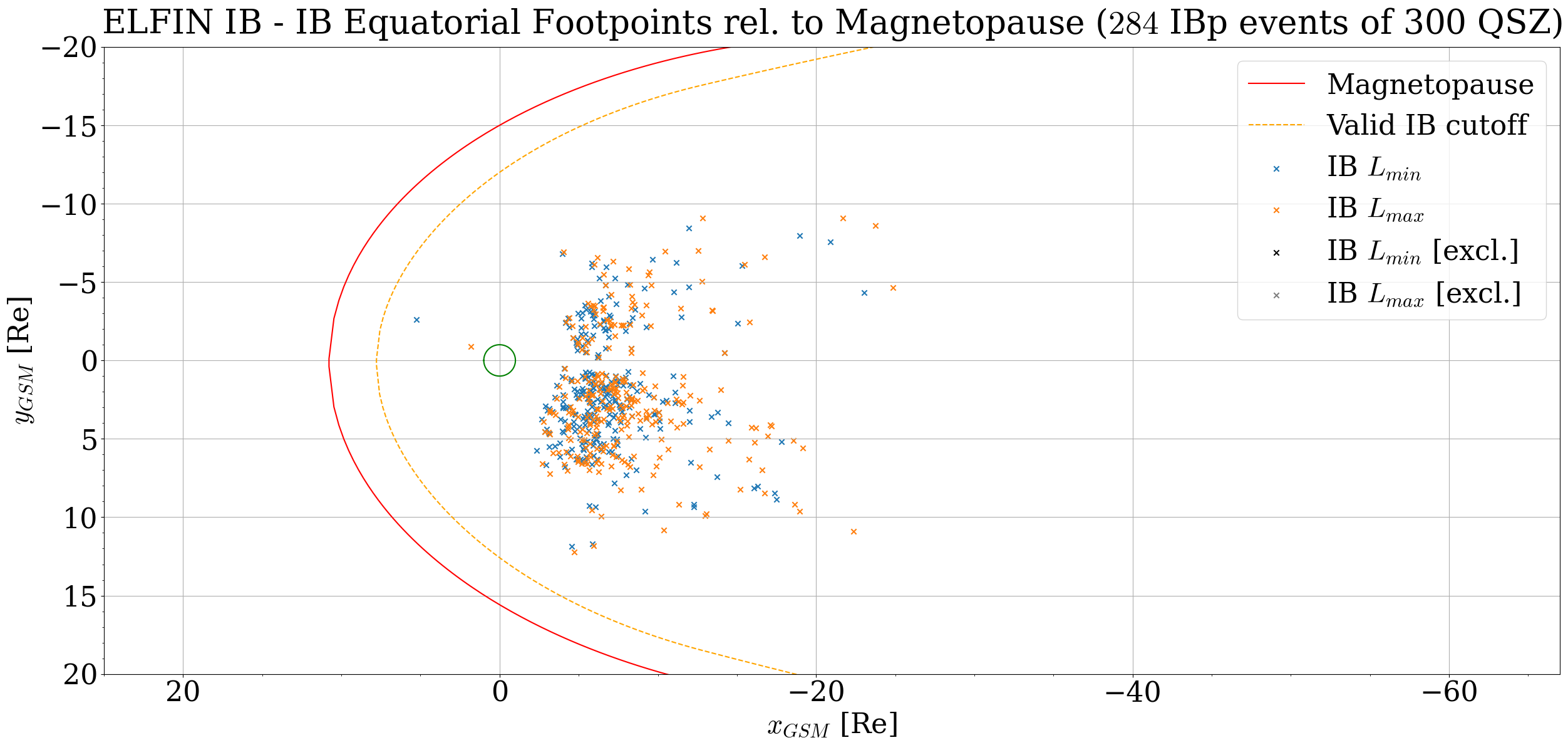}
\caption{Spatial profile scatter plot of approximate locations in the magnetosphere where the equatorial foot-points of electron IB (top) and proton IB (bottom) crossings map, using the T89 field-model. The points labeled $L_{min}$ and $L_{max}$ correspond to the closest and furthest L-shells in each individual IB crossing, nominally corresponding to the highest energy channel present (variable from event to event; see Fig. 5), and to the lowest energy channel present (typically always 63 keV), respectively. It can be seen that electron IBs span a much wider range of equatorial distances from Earth, while proton IBs tend to be more confined to the outer ring current. We note also that the satellite coverage permitted proton observations only on the nightside in the shown MLTs, which would otherwise extend into the flanks and dayside (as with the electrons). We note that for electrons, the mapping is sometimes in the vicinity of or external to the magnetopause. Such events are excluded from our statistics by the same criteria in \citeA{Wilkins2023}.}
\label{Figure 3.}
\end{figure}

\begin{figure}
\begin{centering}
\includegraphics[width=\textwidth]{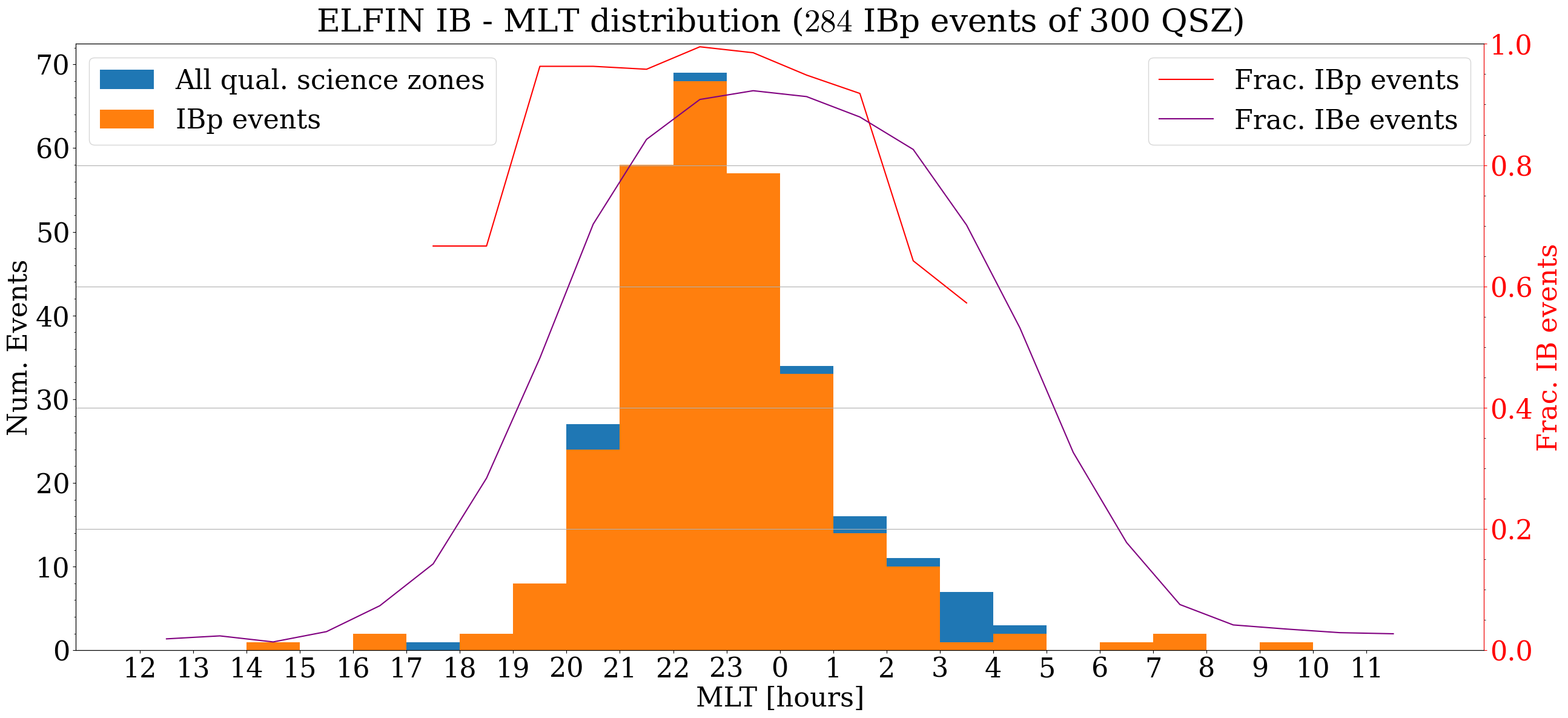}\\
\includegraphics[width=\textwidth]{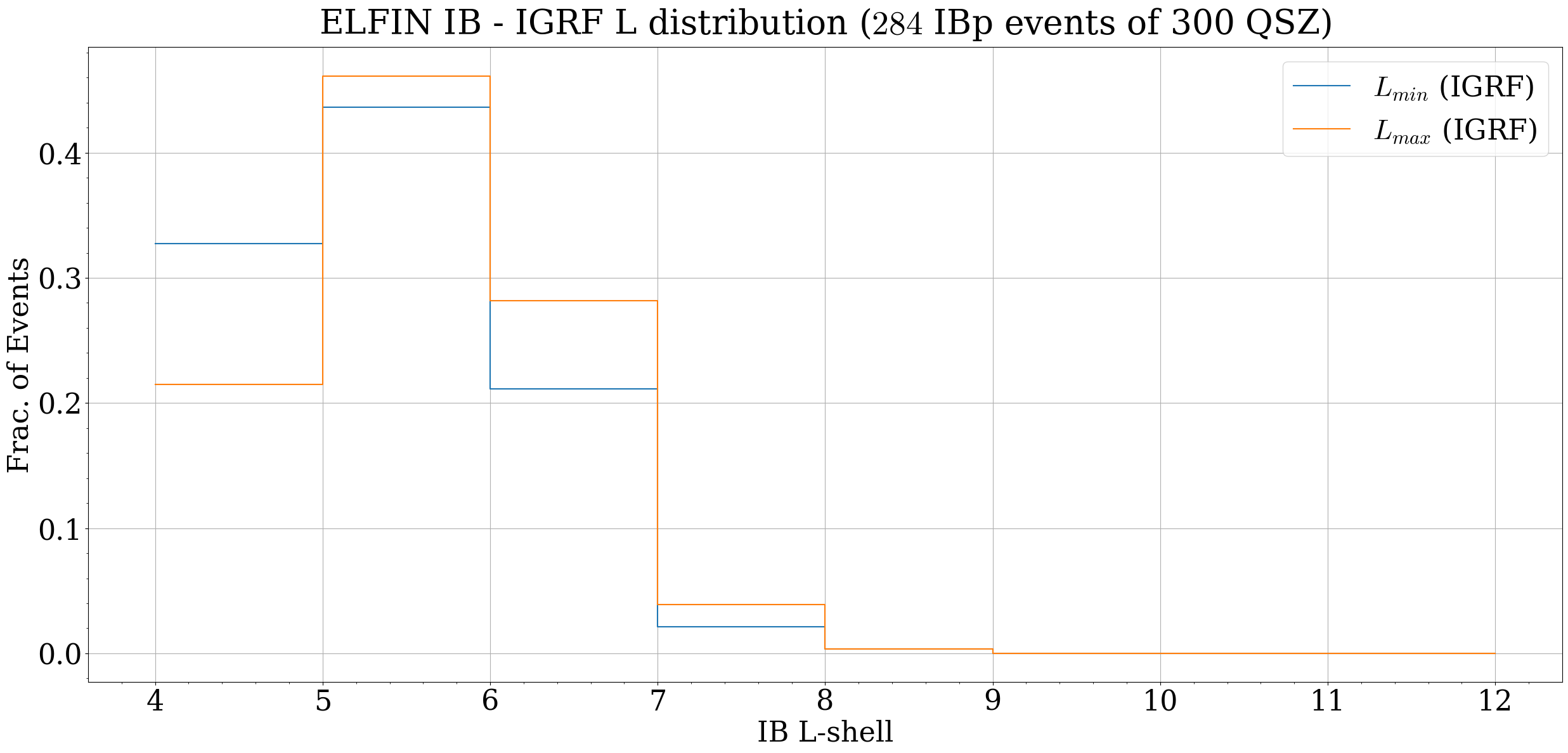}\\
\includegraphics[width=\textwidth]{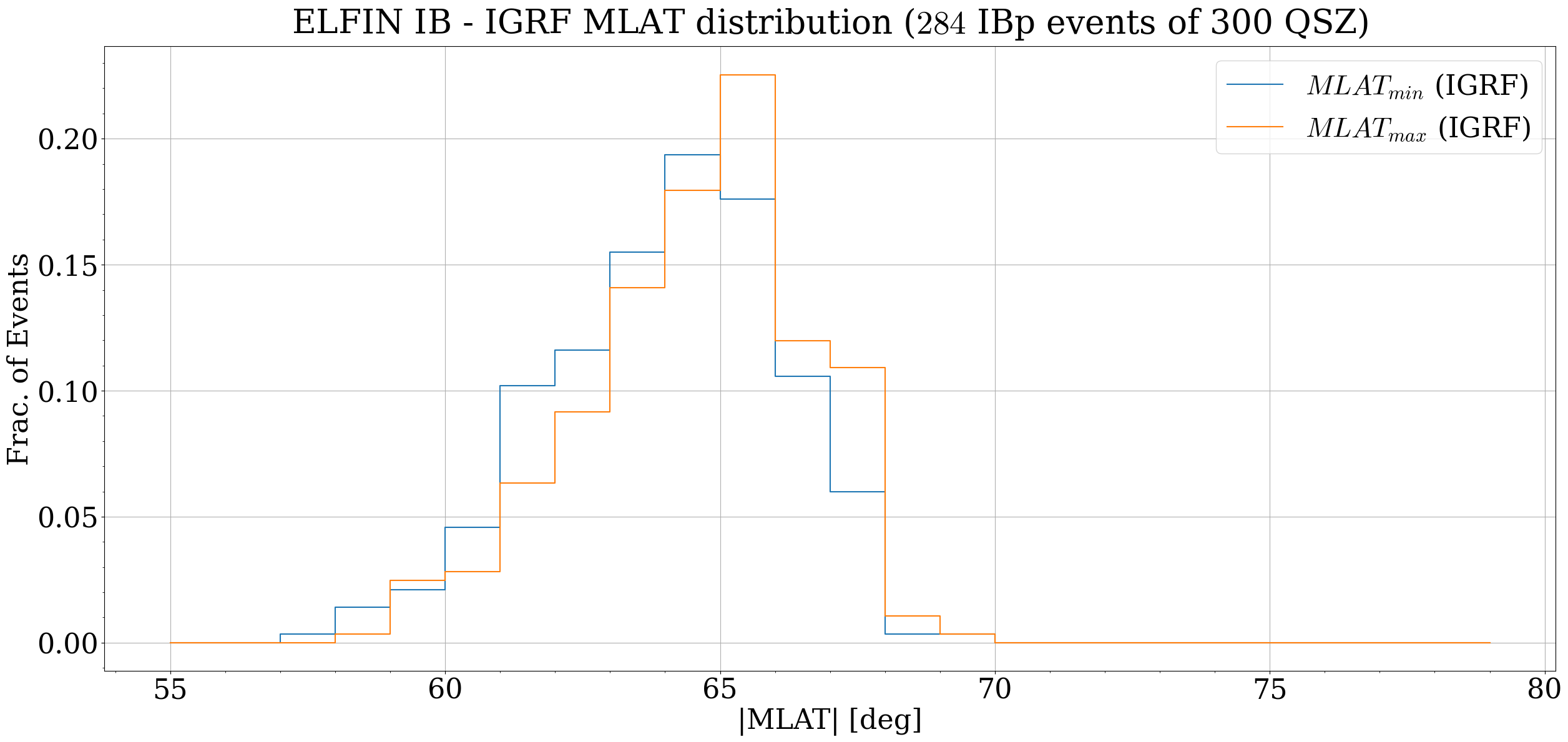}\\
\end{centering}
\caption{Occurrence rate and spatial distribution of proton IBs. Panel (a) shows the distribution of proton IB events versus MLT, with the total number of ELFIN crossings shown in blue, with the overlapping orange bars indicating the total number of proton IB events. The red curve is the occurrence rate for each hourly MLT bin, normalizing for satellite crossing residence. Proton IBs appear with 90-100\% occurrence between 19-03 UT. For comparison the occurrence rate of electron IBs is overplotted in purple, revealing slightly lower occurrence rates, with a peak occurrence occurring 1-2 hours east of the proton peak. Panels (b) and (c) show the spatial distribution of proton IBs in L-shell (IGRF) and MLAT (IGRF) over all local times, with ``min'' and ''max'' quantities akin to Fig. 3. Proton IBs most typically span from L of 5-6, with a sharp drop-off above L=7. The equivalent trend is exhibited for MLAT, where a peak occurrence of 64$^\circ$-66$^\circ$ is observed, with a sharp fall beyond 68$^\circ$ and below 61$^\circ$.}
\label{Figure 4.}
\end{figure}

\begin{figure}
\begin{centering}
\noindent\includegraphics[width=\textwidth]{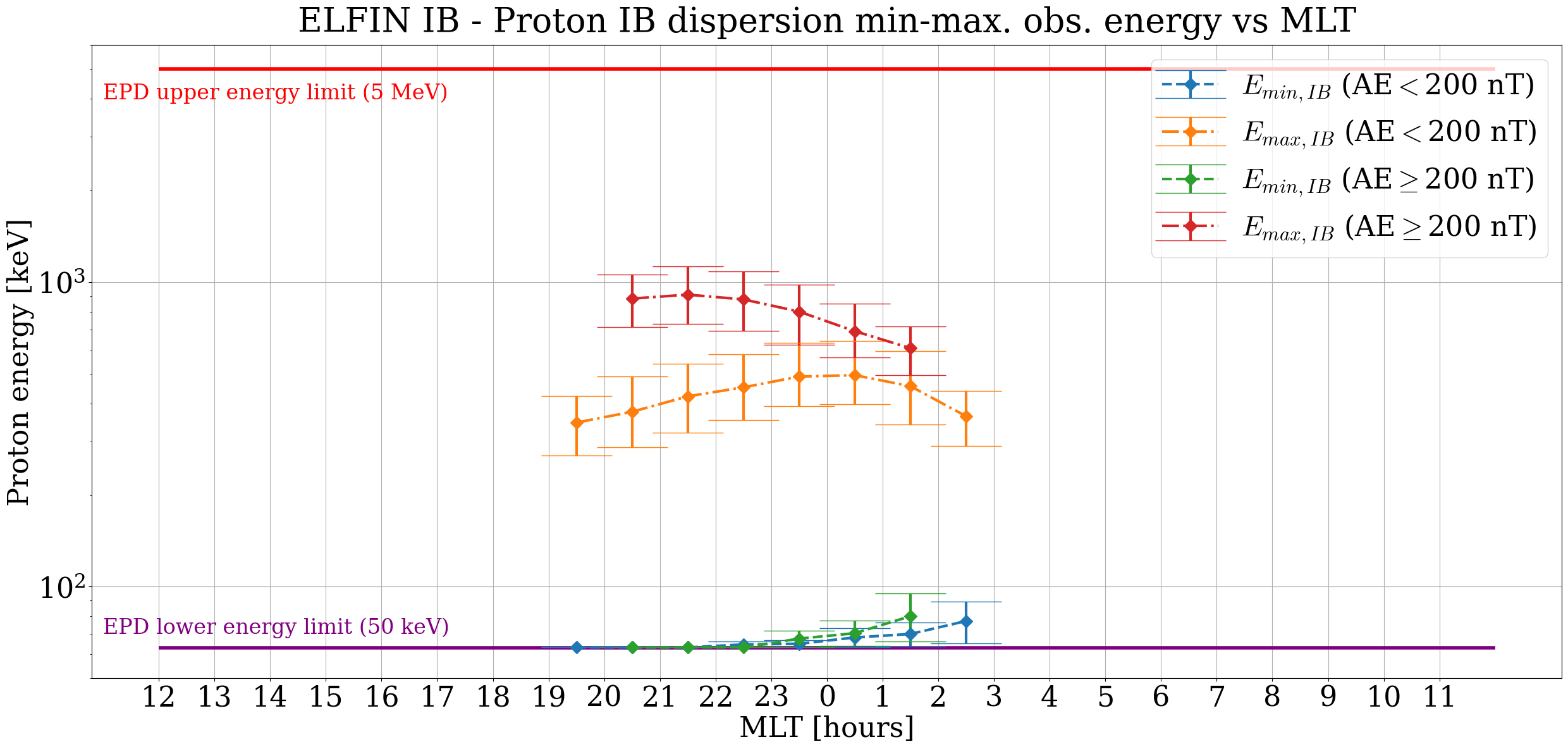}\\
\noindent\includegraphics[width=0.5\textwidth]{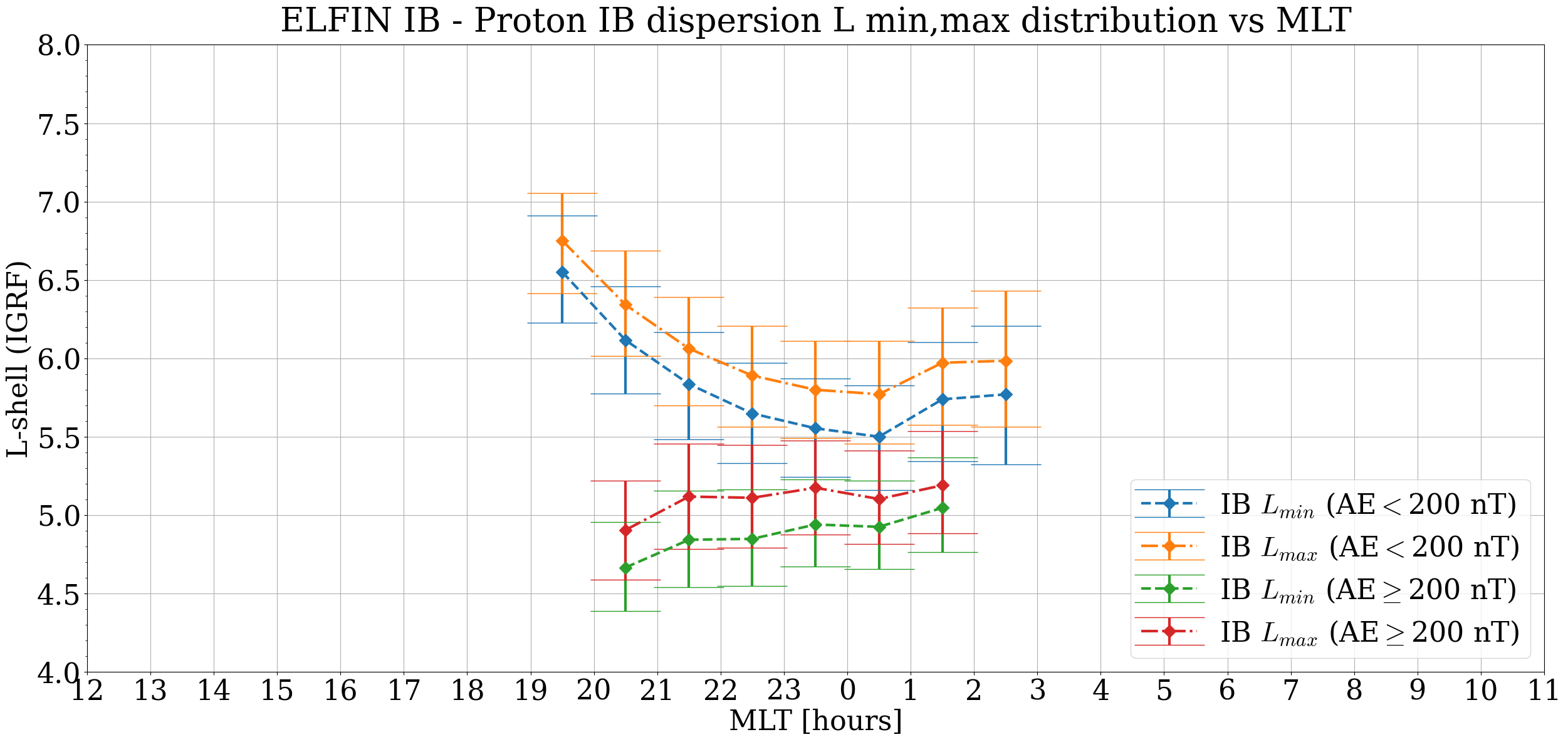}\includegraphics[width=0.5\textwidth]{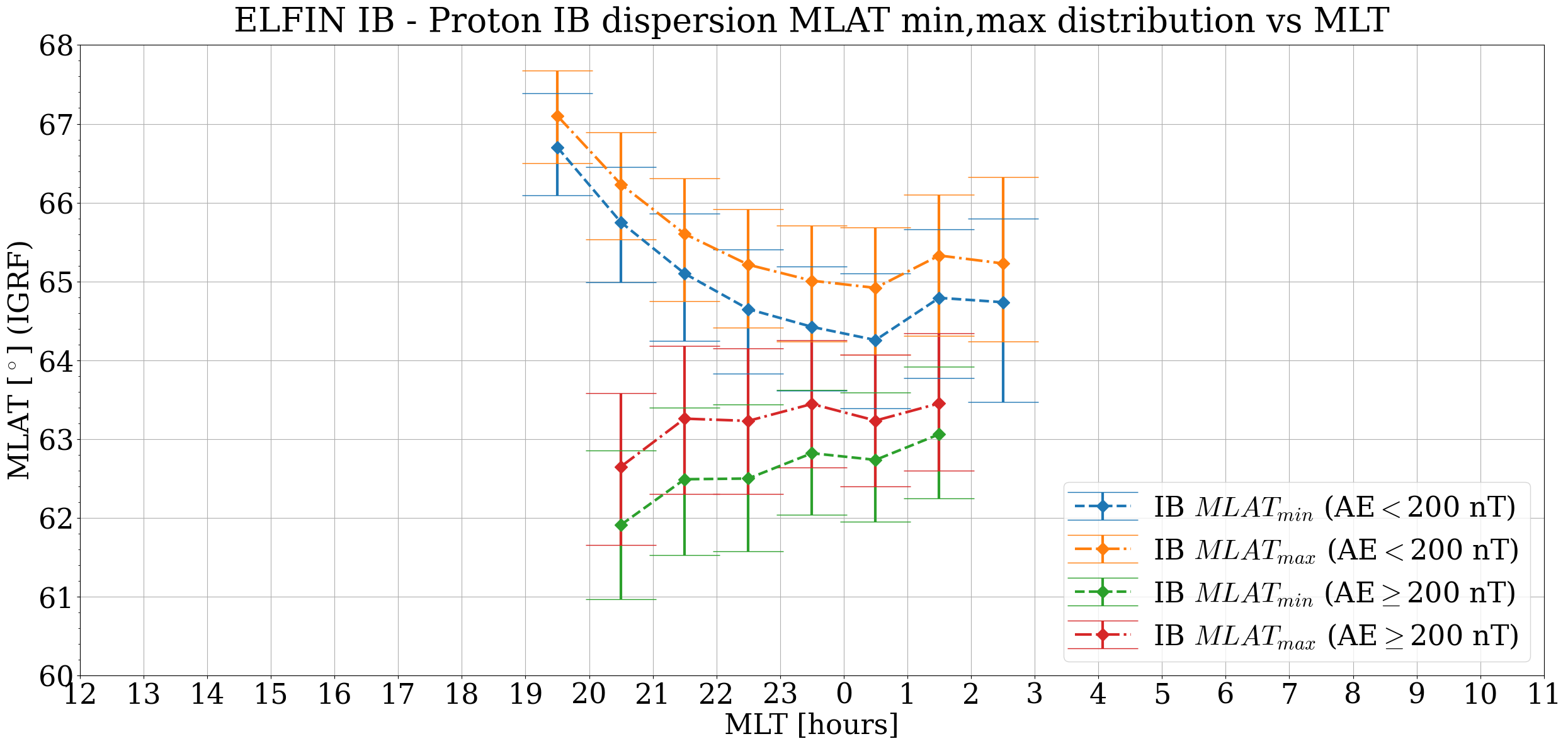}\\
\noindent\includegraphics[width=0.5\textwidth]{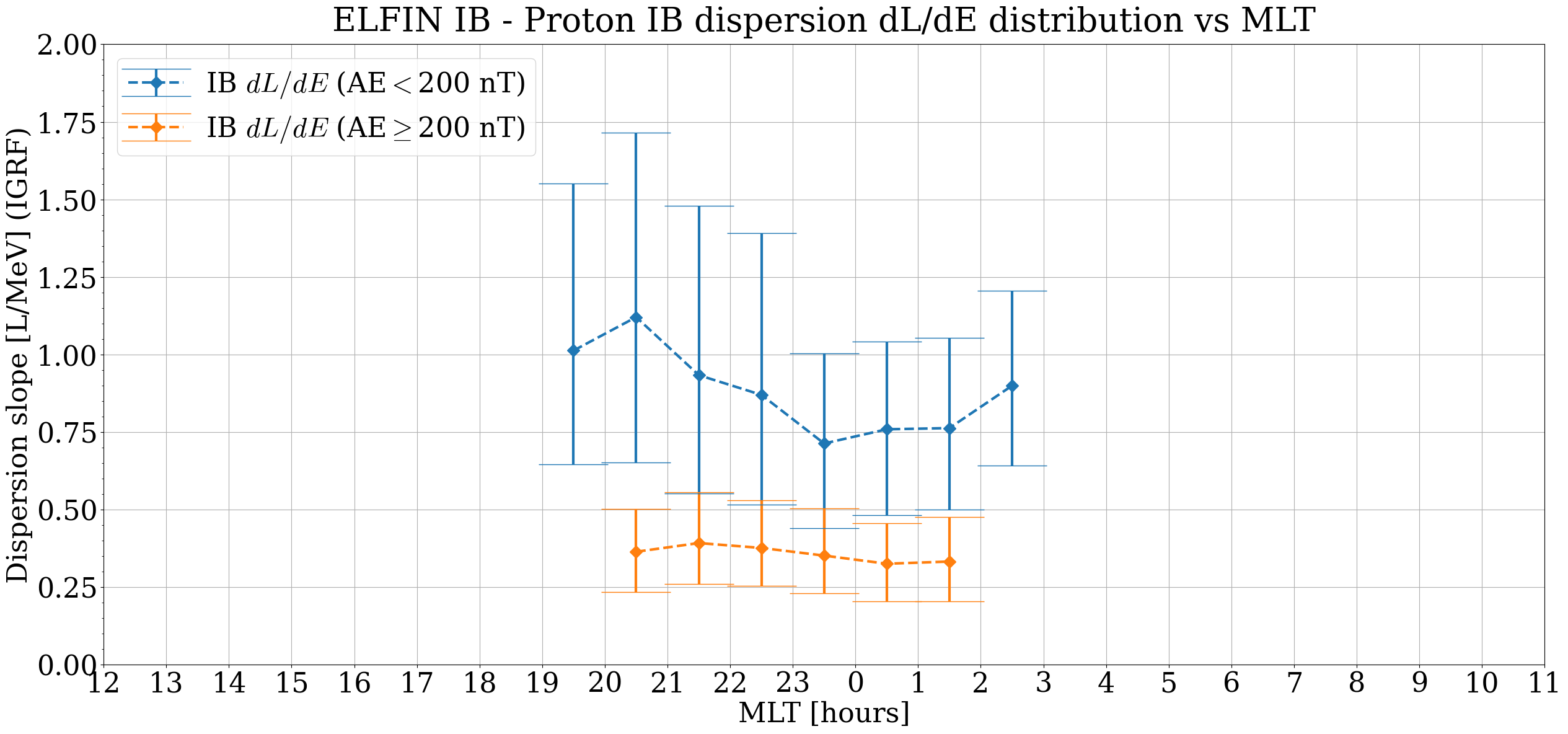}\includegraphics[width=0.5\textwidth]{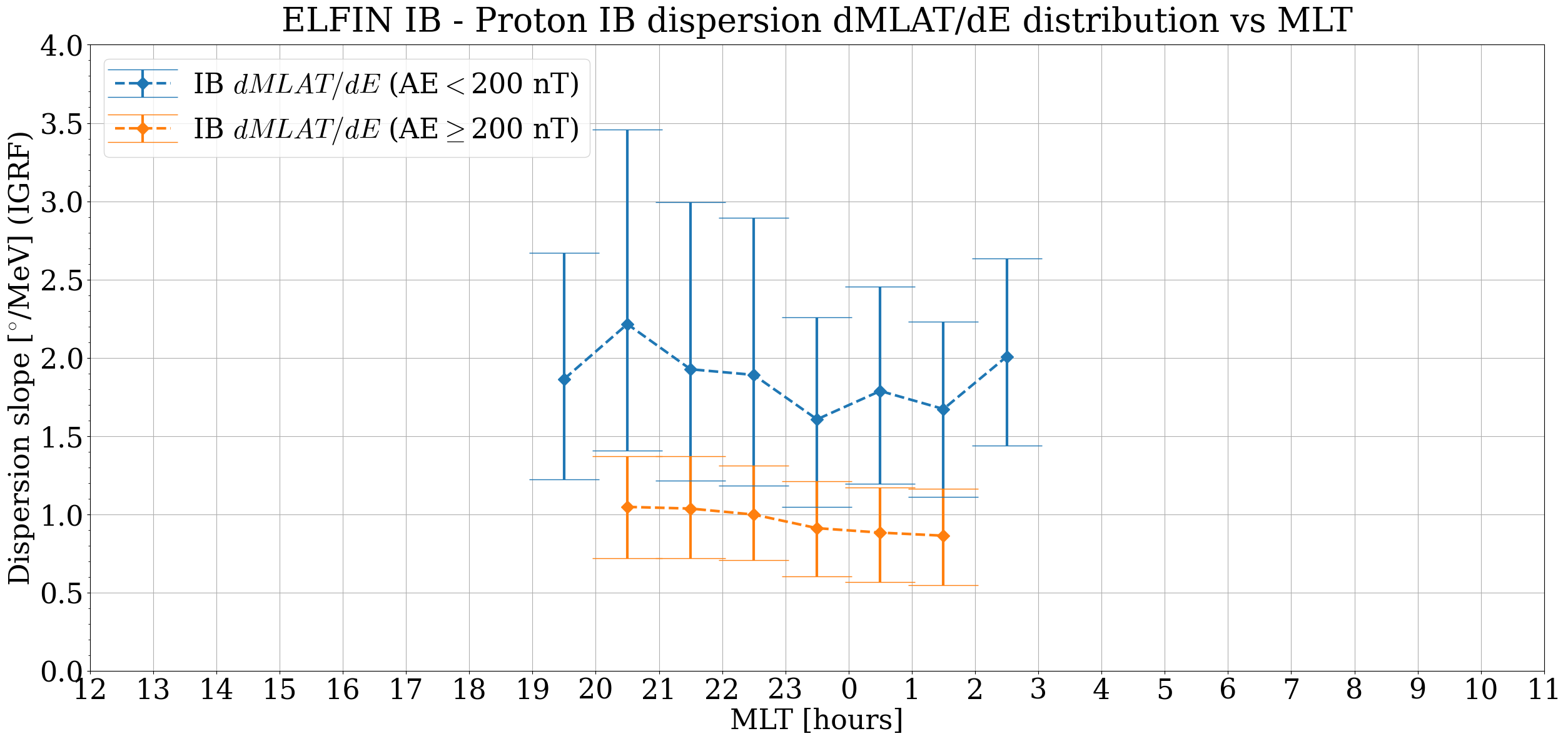}\\
\end{centering}
\caption{MLT distribution of proton IB characteristic means and standard deviations versus activity. Panel (a) shows the minimum and maximum energies (bounding a continuous energy spectrum) appearing in the IB crossings, with peak energies increasing by up to 500 keV between quiet and non-quiet intervals, shifting toward pre-midnight. The lowest energy is typically always the lowest ELFIN can resolve, suggesting the IBs extend to lower energies. Panels (b) and (c) show the L-shells and MLAT versus activity, which exhibit a symmetric U-shape distribution at quiet time, which is disturbed at active time in the pre-midnight sectors. Panels (d) and (e) show the energy-latitude dispersion slopes for these cases.}
\label{Figure 5.}
\end{figure}

\begin{figure}
\begin{centering}
\noindent\includegraphics[width=\textwidth]{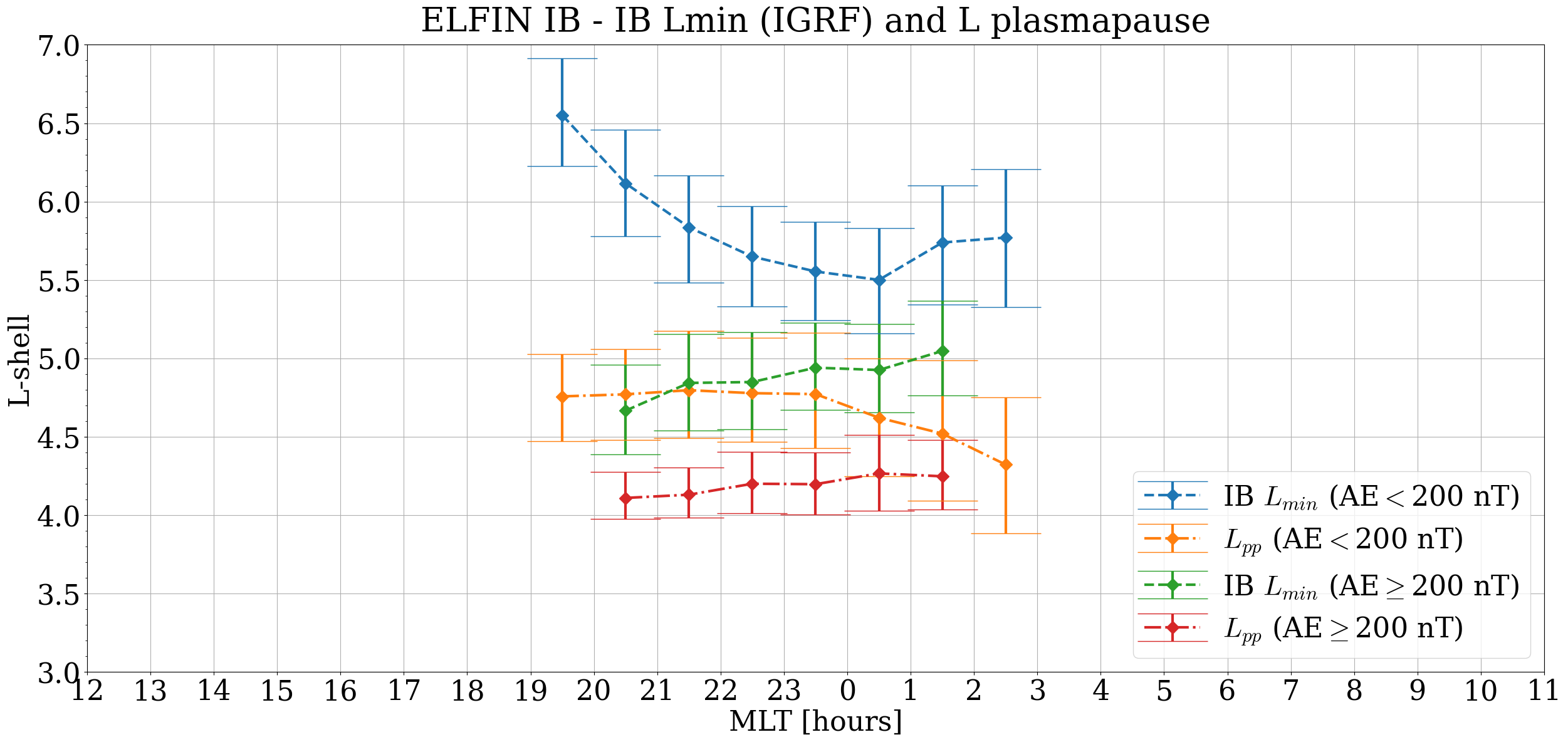}\\
\noindent\includegraphics[width=\textwidth]{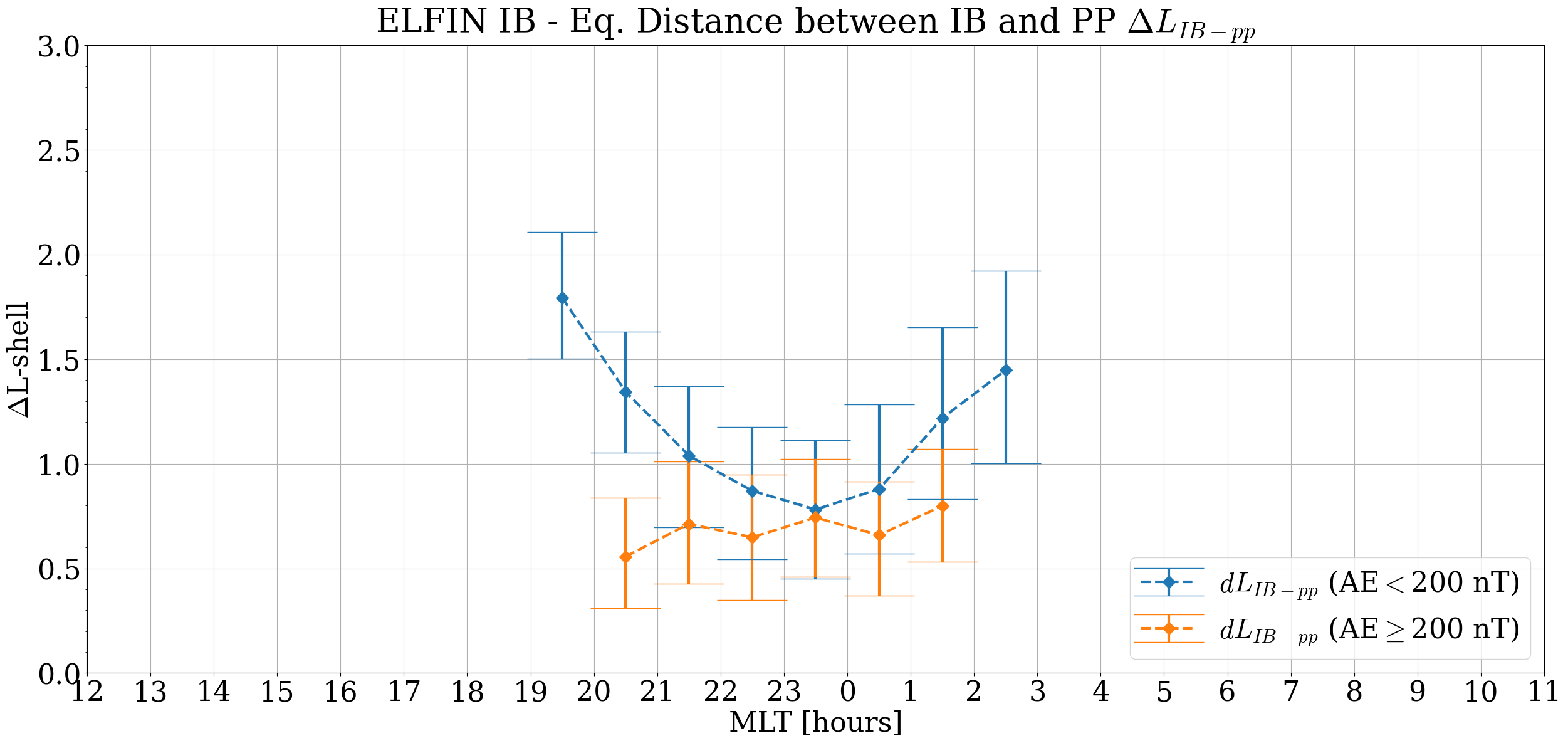}
\end{centering}
\caption{Comparison of plasmapause location with proton IBs versus MLT and activity, based on the model of \citeA{OBrien2003}. The minimum IB locations (by definition closest to the plasmapause; typically highest-energy IB portion) are plotted for quiet and active intervals, where it can be seen that they are generally separated by 1-2 Re at quiet time, reducing to 0.5-1 Re at active time. The bottom Panel considered the event-wise different $\Delta L_{IB-pp} = L_{min,IB}-L_{pp}$, revealing the same trend. }
\label{Figure 6.}
\end{figure}

\begin{figure}
\begin{centering}
\noindent\includegraphics[width=\textwidth]{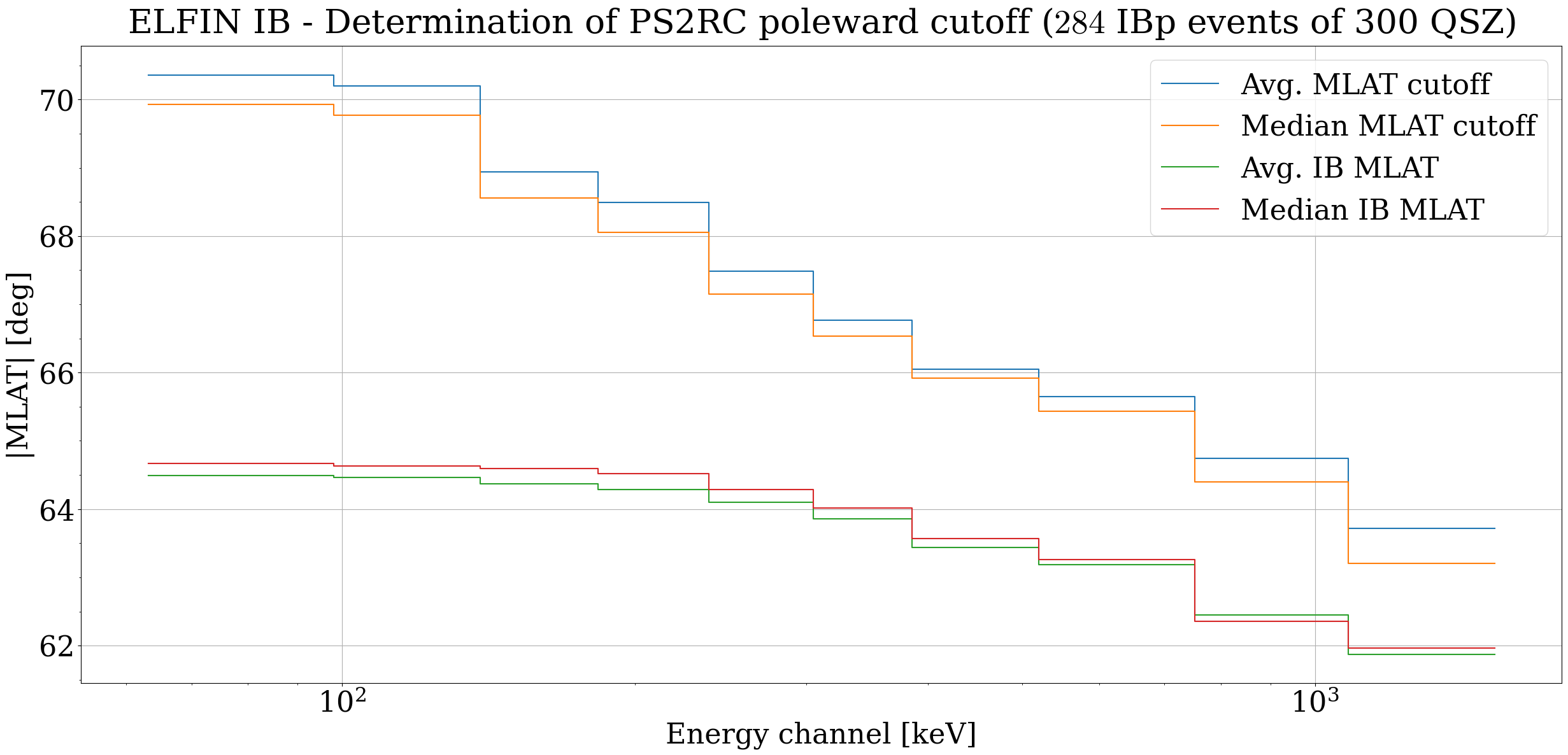}\\
\noindent\includegraphics[width=\textwidth]{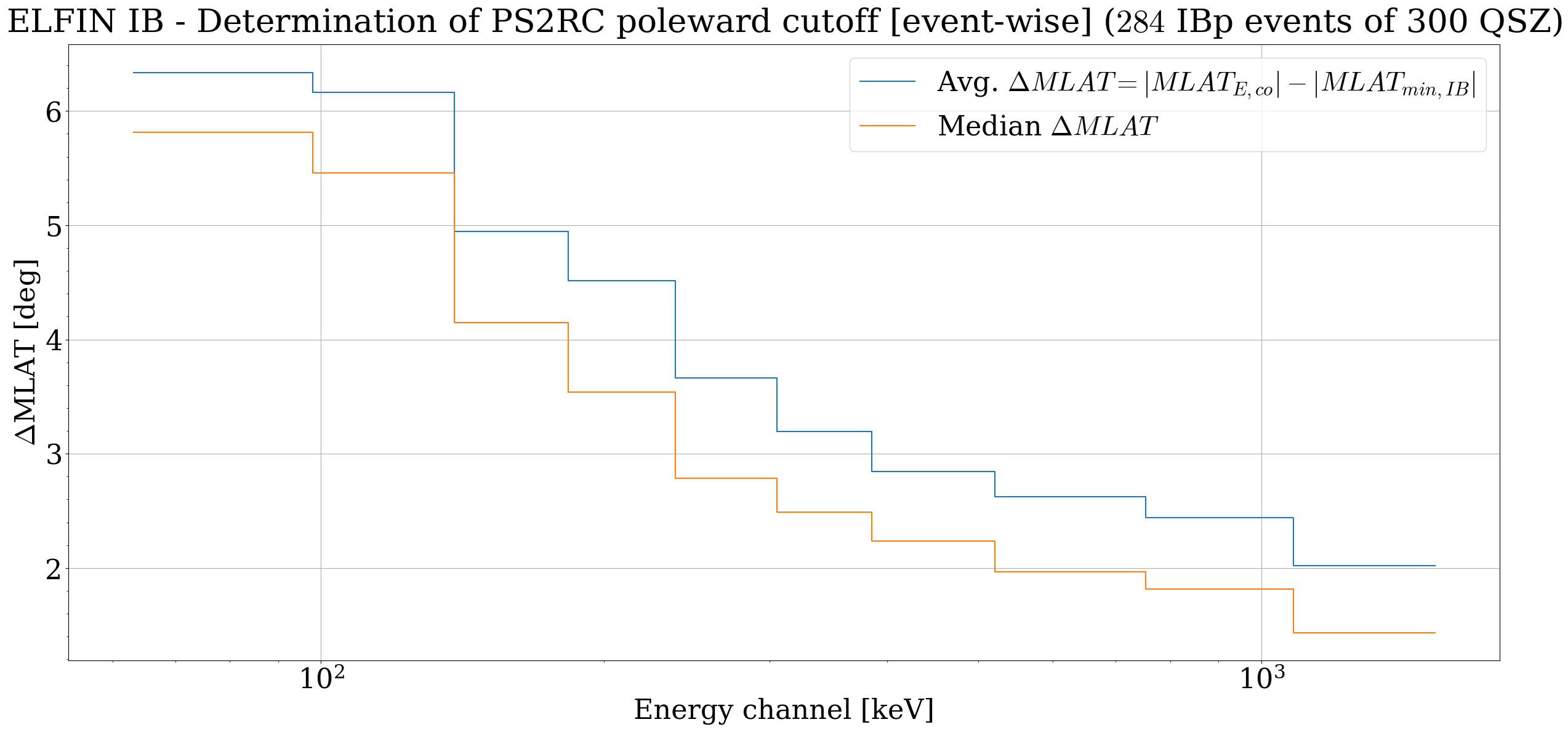}\\
\noindent\includegraphics[width=\textwidth]{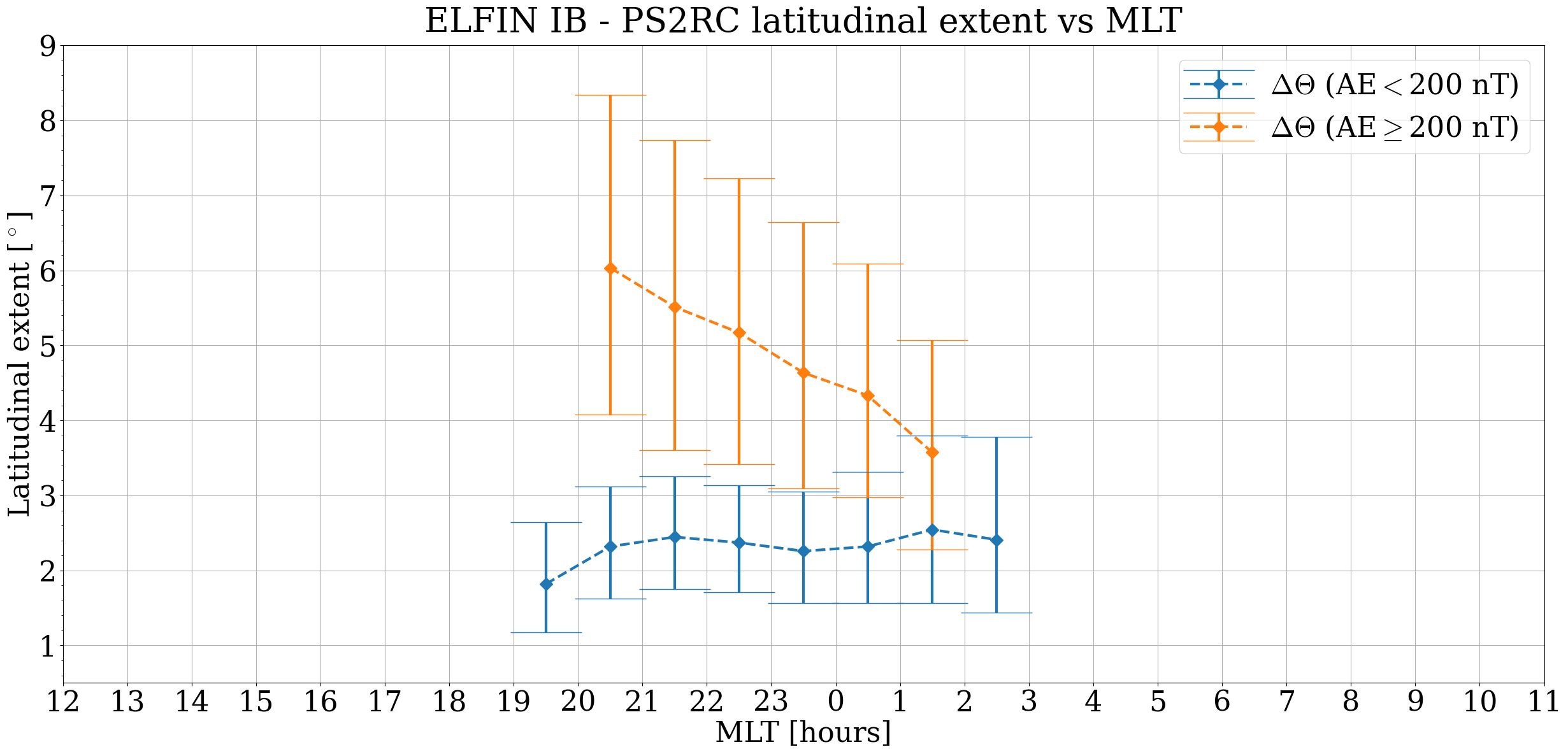}
\end{centering}
\caption{Determination of the plasma sheet inner edge in ELFIN IB events. The top Panel shows the absolute MLAT over all IB events by E channel for which an IB was observed. A rapid rise in slope of omni-directional flux dropout is observed at energies 200 keV and below, suggesting the appearance of a plasma sheet population separate from the FLC-dominated PS2RC. The middle Panel shows the event-wise difference, revealing a similar increase below 300 keV. The bottom Panel shows the latitudinal width $\Delta \Theta$ of the curvature scattering region (PS2RC) versus MLT and activity based on the 300 keV drop-out criterion. The latitudinal extent is relatively uniform at quiet time across MLT, but exhibits significant asymmetry at active time, growing in the pre-midnight sector.}
\label{Figure 7.}
\end{figure}

\begin{figure}
\noindent\includegraphics[width=\textwidth]{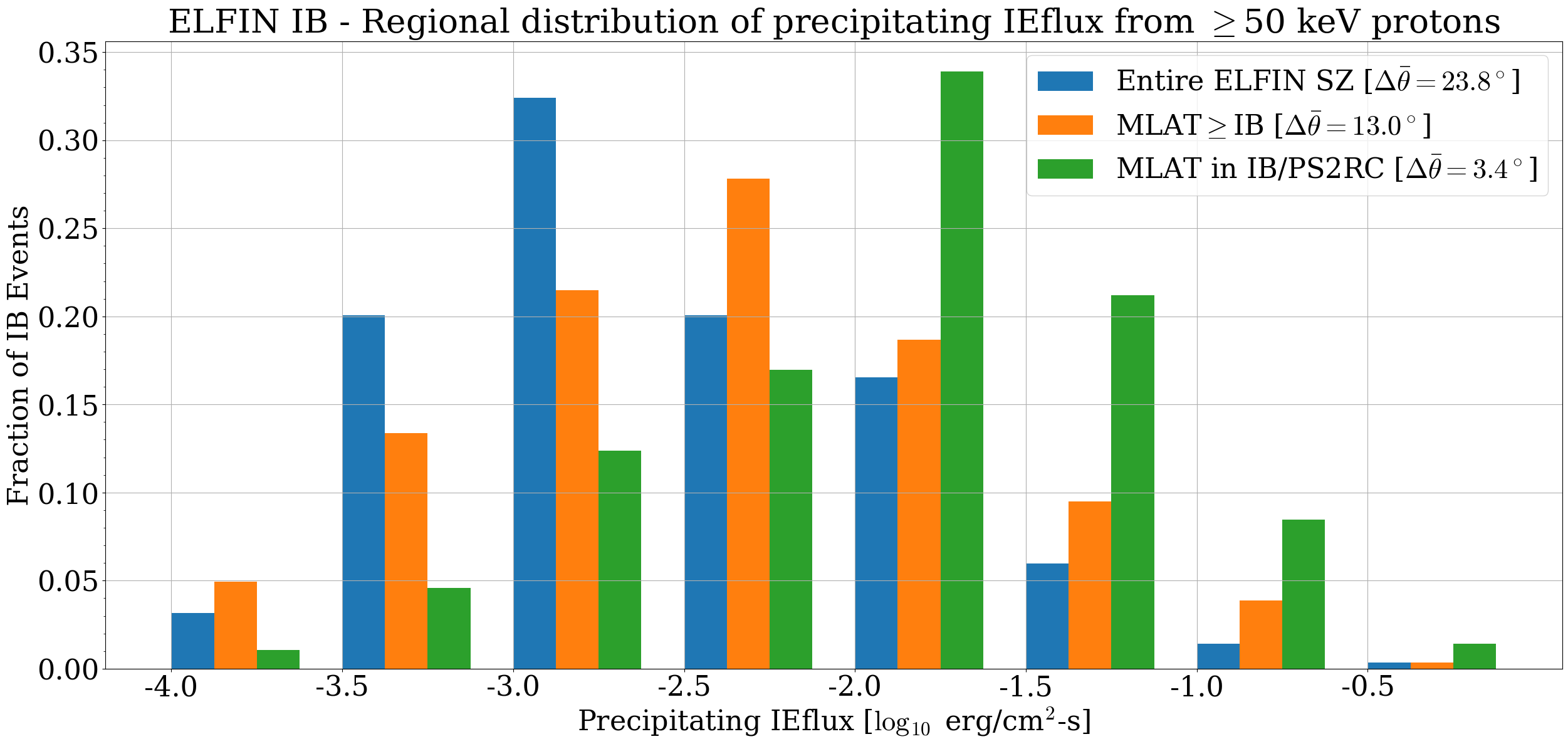}
\caption{Distribution of latitude-averaged integral precipitating energy flux (IEflux) from $>$50 keV protons in different latitudinal ranges of the ELFIN IB event dataset, alongside their average latitudinal extent ($\Delta \bar{\theta}$). The blue bars represent the entire ELFIN science zone (nominally 55$^\circ$ to 80$^\circ$ latitude), including ring current, outer radiation belt, plasma sheet, and partial polar cap. Orange bars represent all latitudes poleward of IBs, including the PS2RC region, but not terminating at its poleward boundary, and thus including the whole plasma sheet. Green bars represent fluxes within the IB and PS2RC interface, which can be seen to by far provide the highest average precipitation, as well as be the most narrow in latitude.}
\label{Figure 8.}
\end{figure}

\begin{figure}
\begin{centering}
\noindent\includegraphics[width=\textwidth]{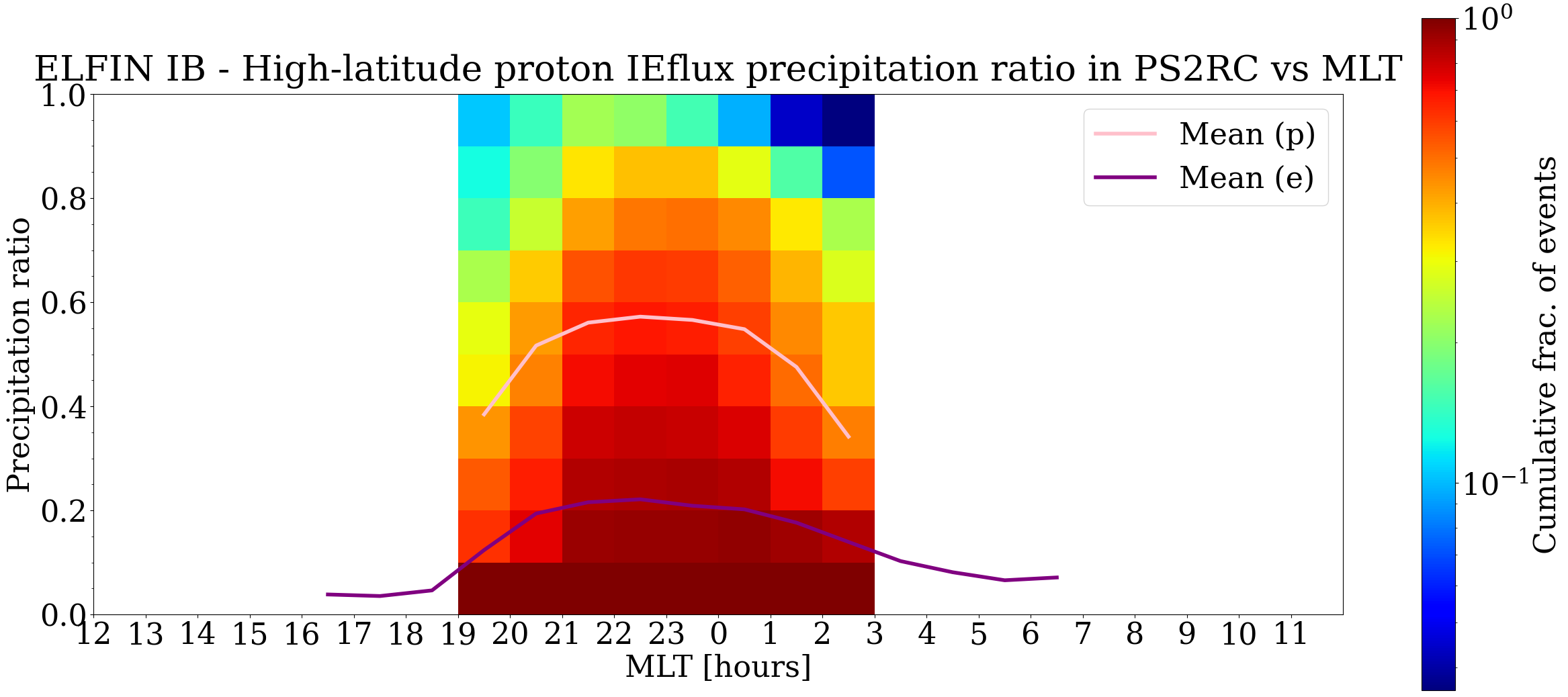}\\
\noindent\includegraphics[width=\textwidth]{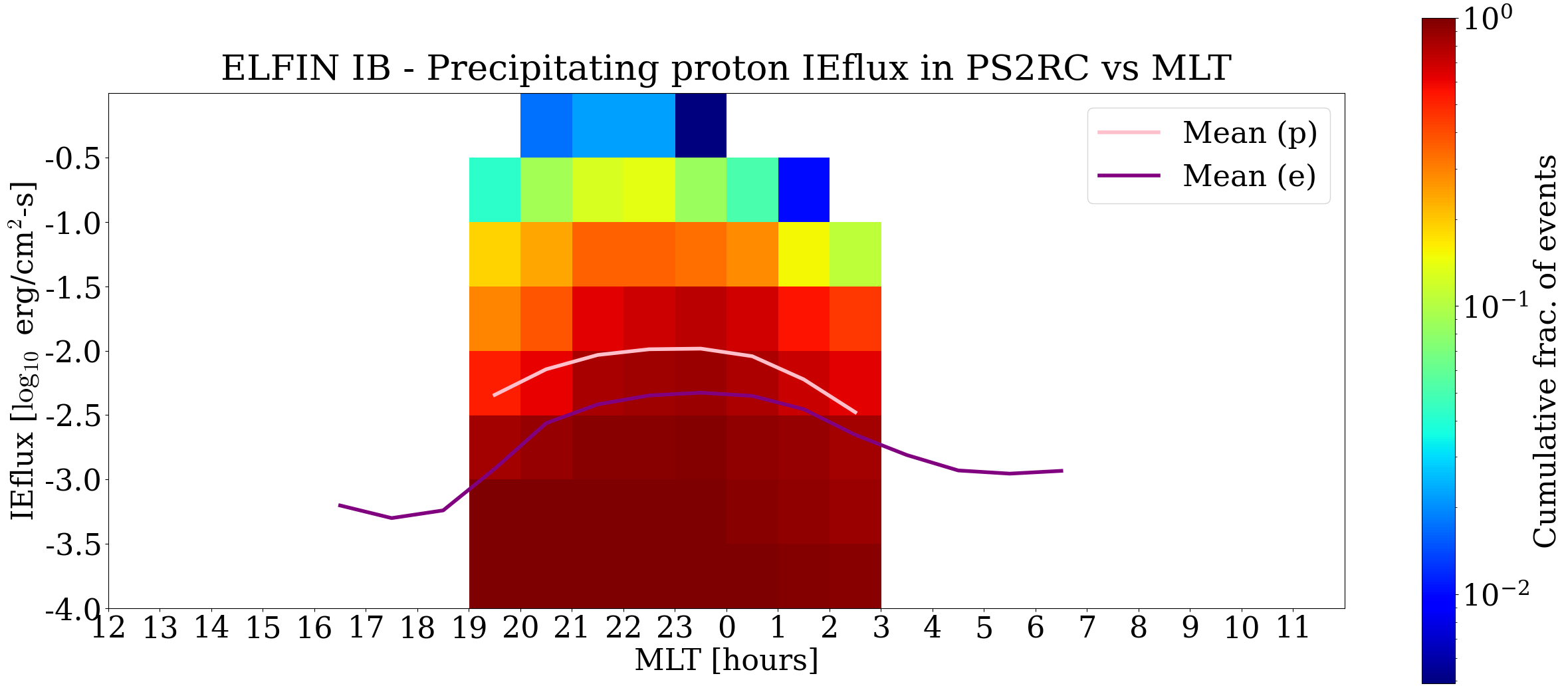}\\
\end{centering}
\caption{Cumulative distribution in MLT of $>$50 keV isotropic proton integral precipitating energy flux (IEflux) ratios (top) and the IEfluxes (bottom). The ratio is computed by integrating the proton precipitation in the PS2RC, and dividing by the total for the entire science zone (55$^\circ$-80$^\circ$). Protons IBs have a highly-symmetric inverted U-shape distribution about 22-23 MLT peaking with $\sim$60\% of the total precipitation on average, trending downward toward dawn and dusk, suggesting FLC is generally the dominant precipitation mechanism for $>$50 keV high-latitude protons throughout much of the nightside. Equivalent electron IB mean values are over-plotted in purple (normalized by their own $>$50 keV precipitation separately from protons), demonstrating that on average FLC scattering is a far more dominant precipitation mechanism for protons than for electrons. The bottom Panel reveals the trends in latitude-averaged precipitating energy-flux within the PS2RC (protons) and PS2ORB (electrons), respectively, with protons precipitating $\sim$3 times as much as electrons on average, at times reaching 1 erg/cm$^2$-s mean over its entire latitudinal range.}
\label{Figure 9.}
\end{figure}

\begin{figure}
\begin{centering}
\includegraphics[width=0.45\textwidth]{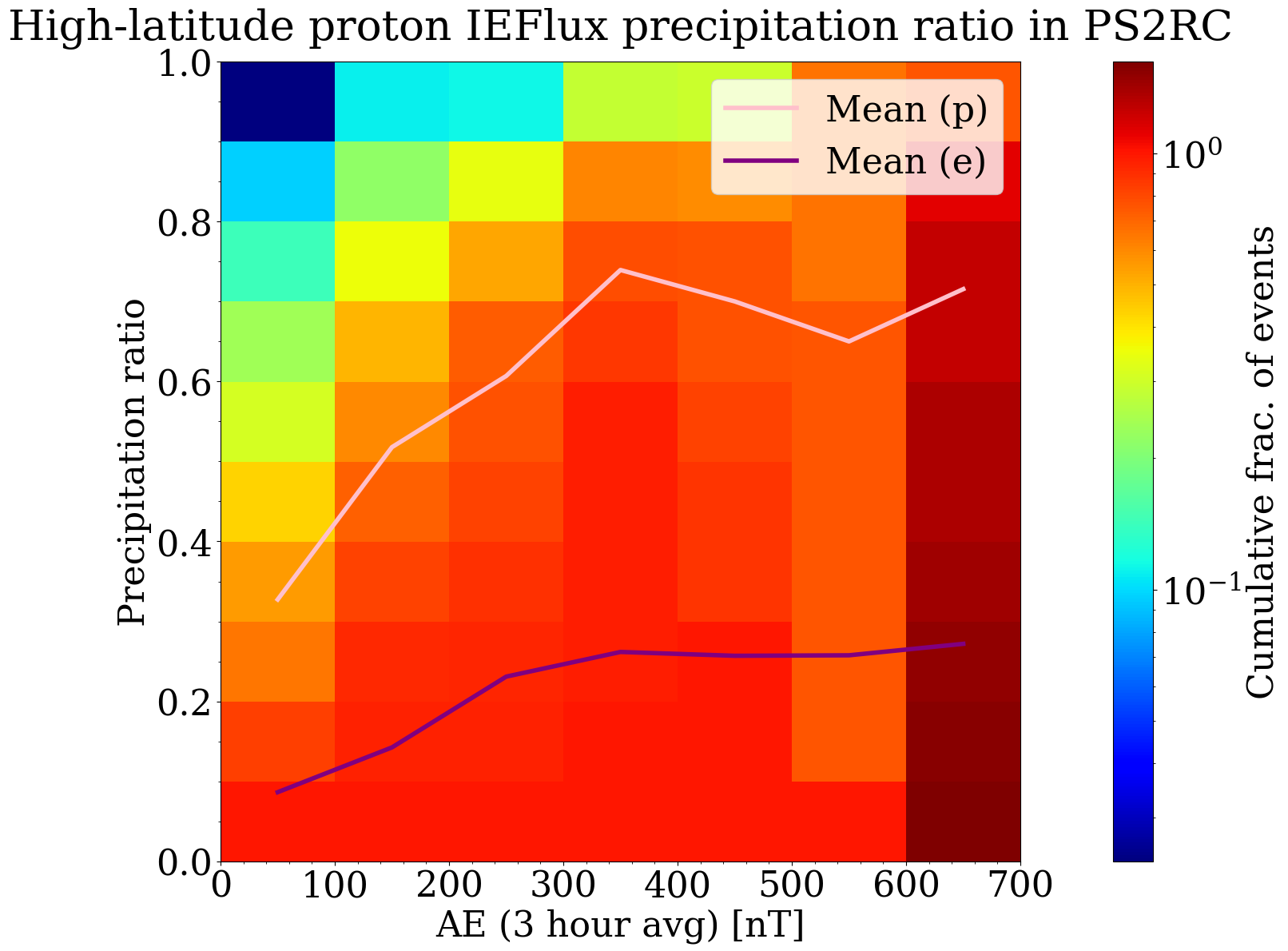}
\includegraphics[width=0.45\textwidth]{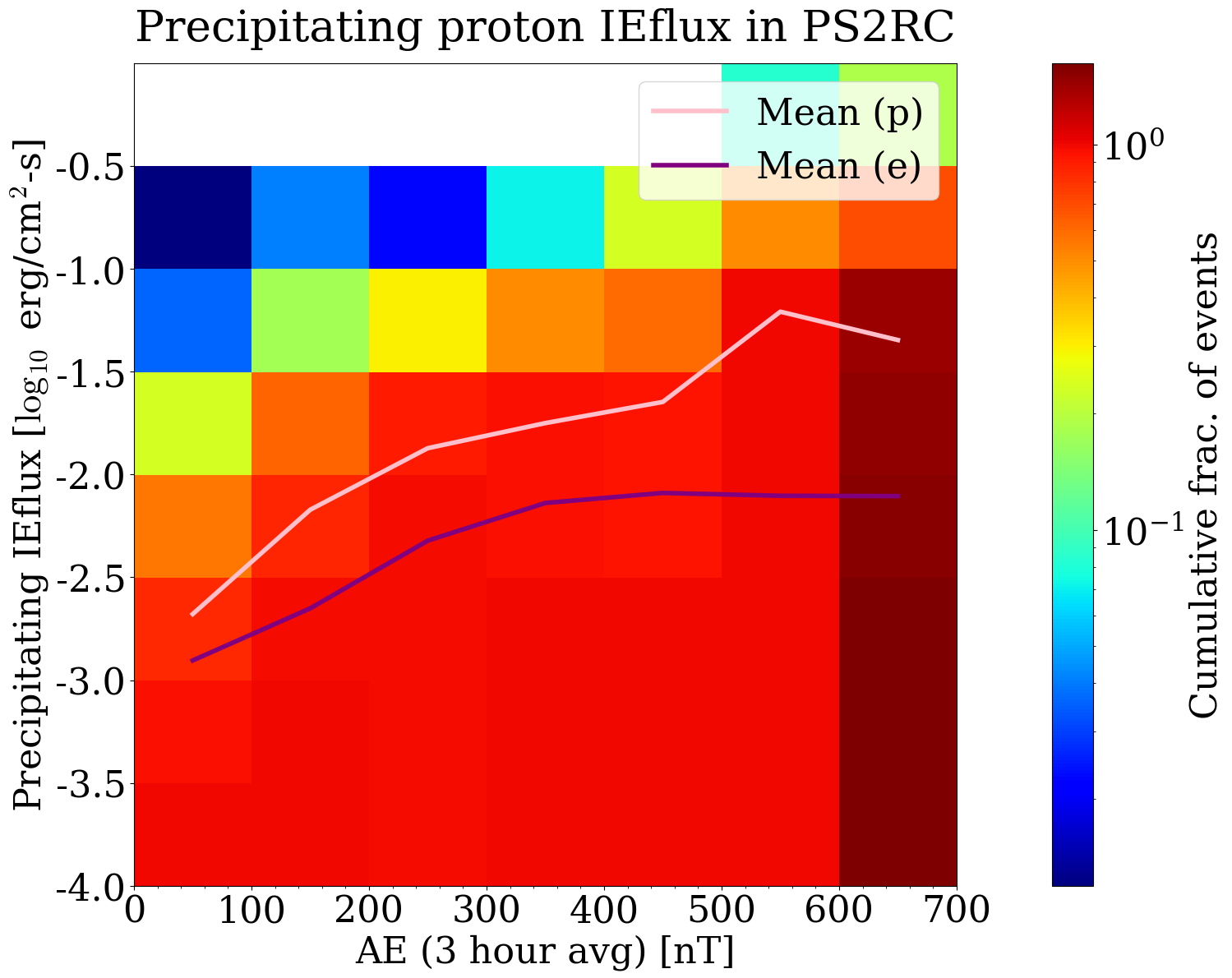}\\
\includegraphics[width=0.45\textwidth]{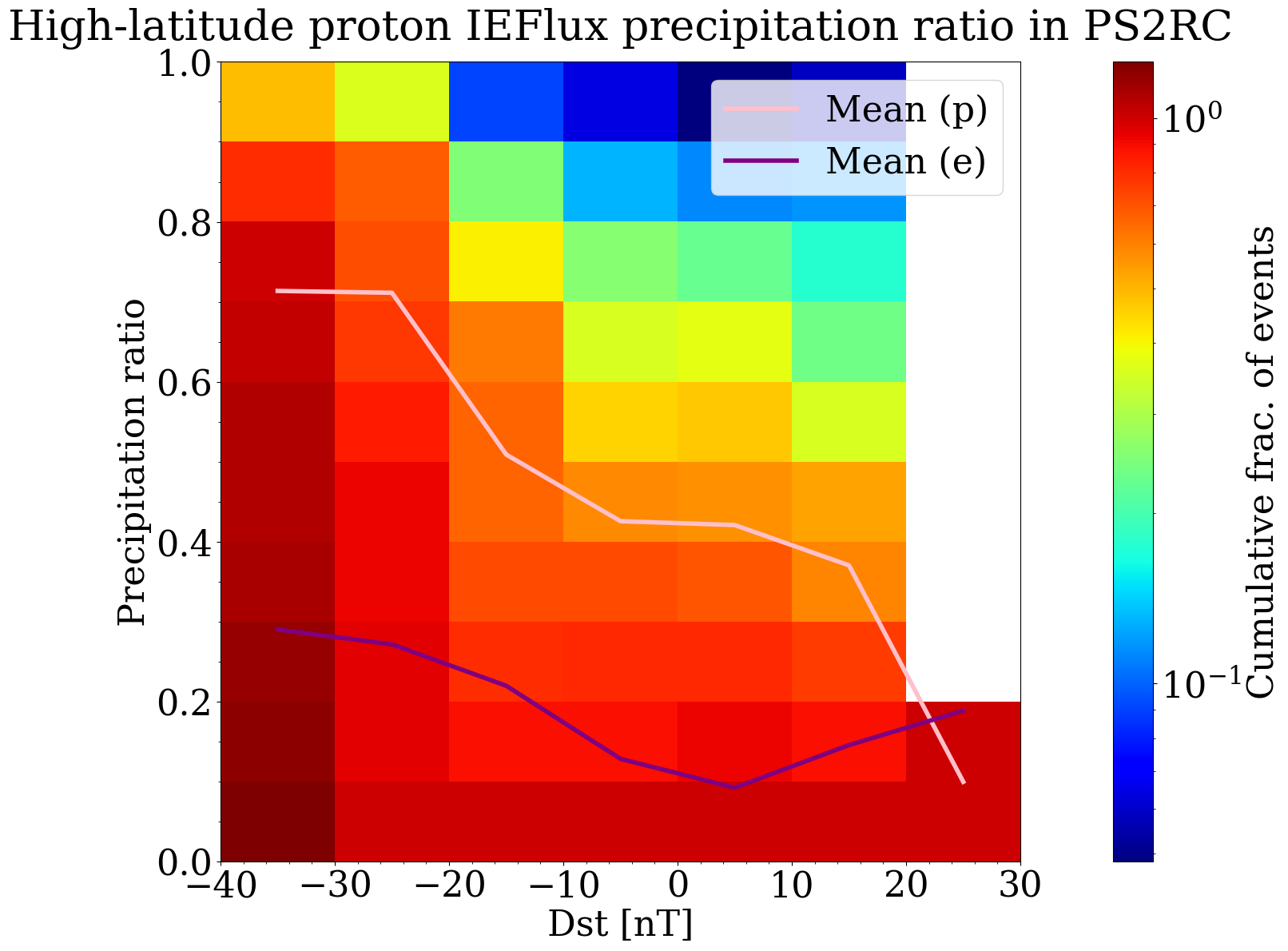}
\includegraphics[width=0.45\textwidth]{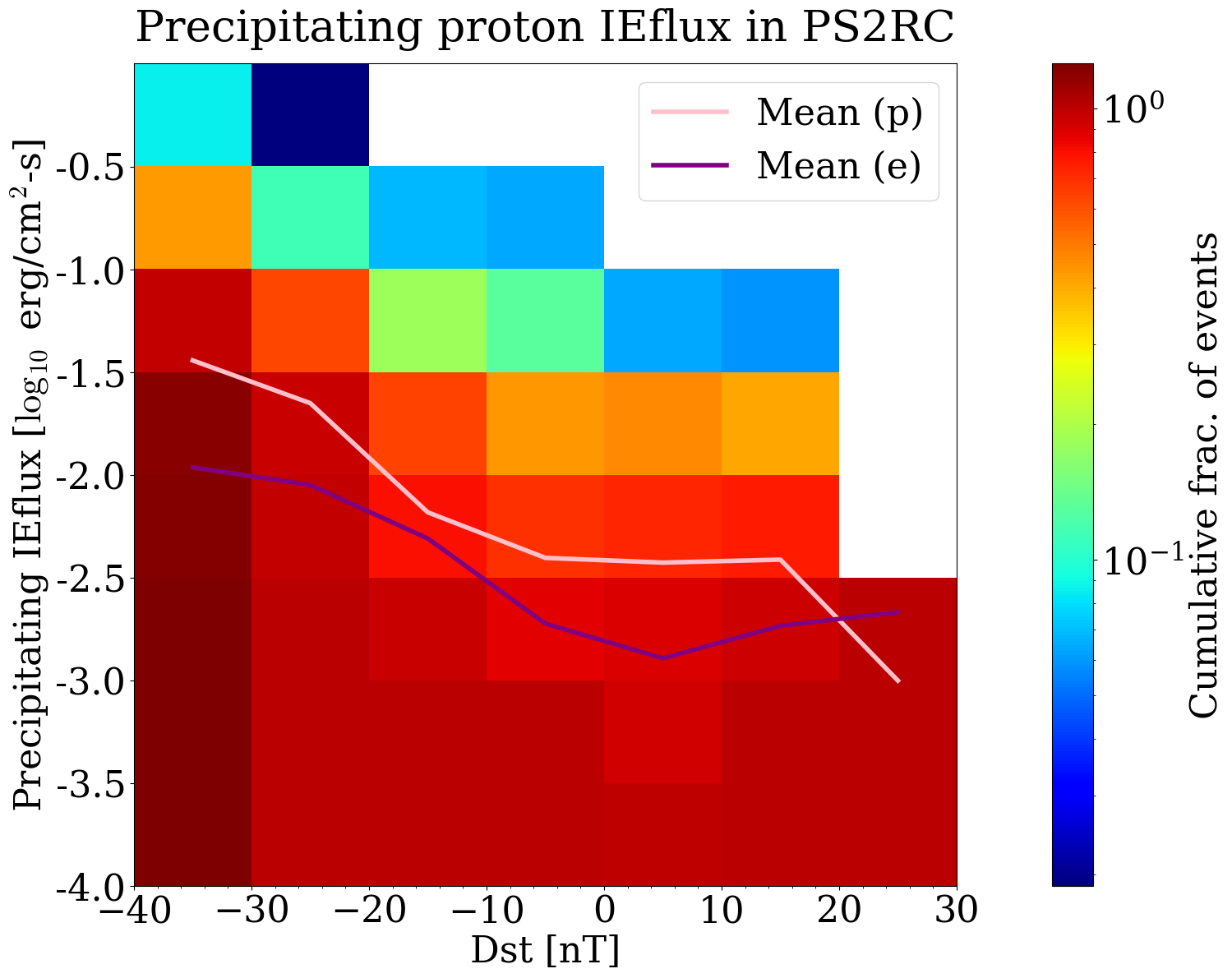}\\
\includegraphics[width=0.45\textwidth]{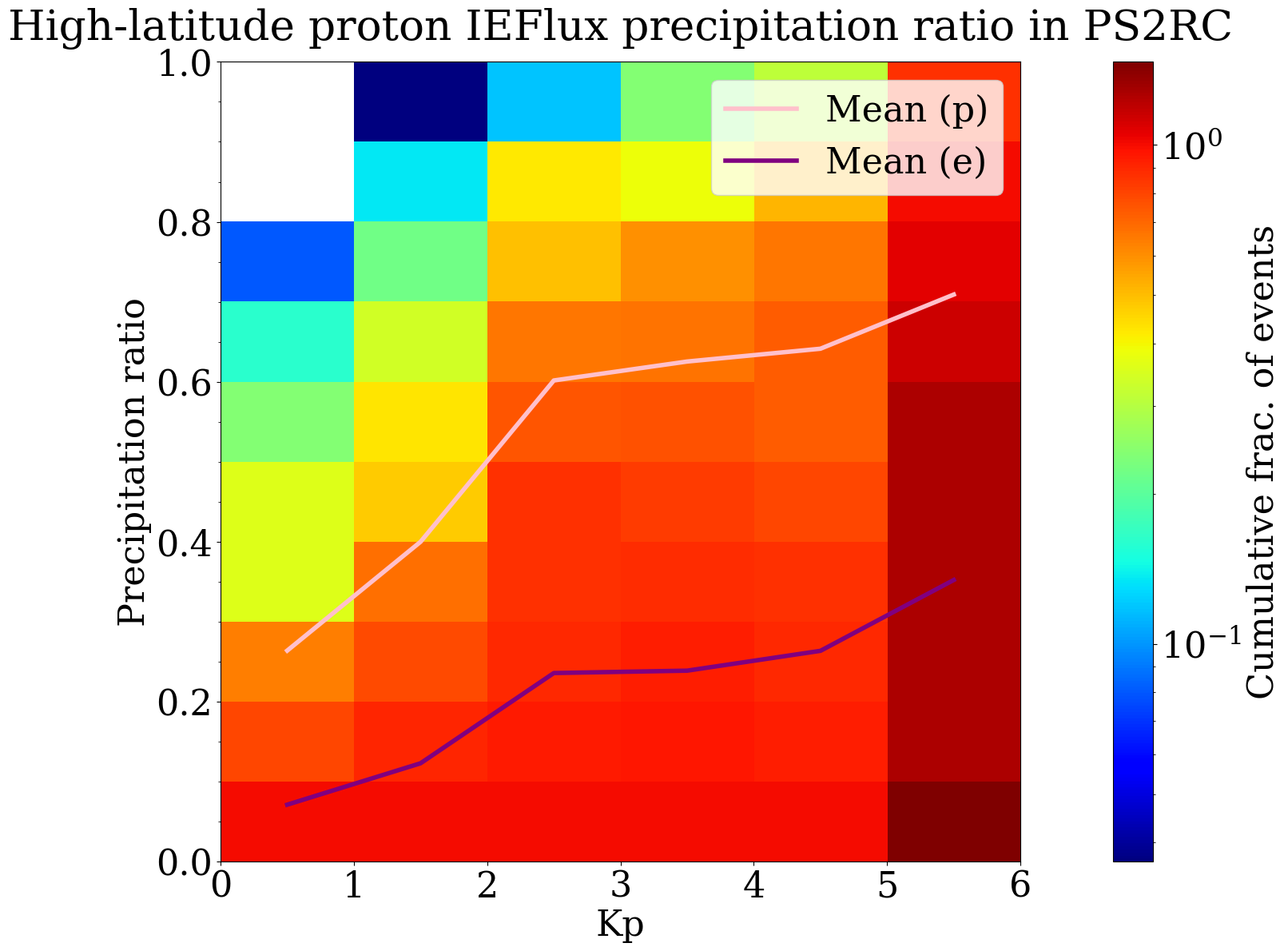}
\includegraphics[width=0.45\textwidth]{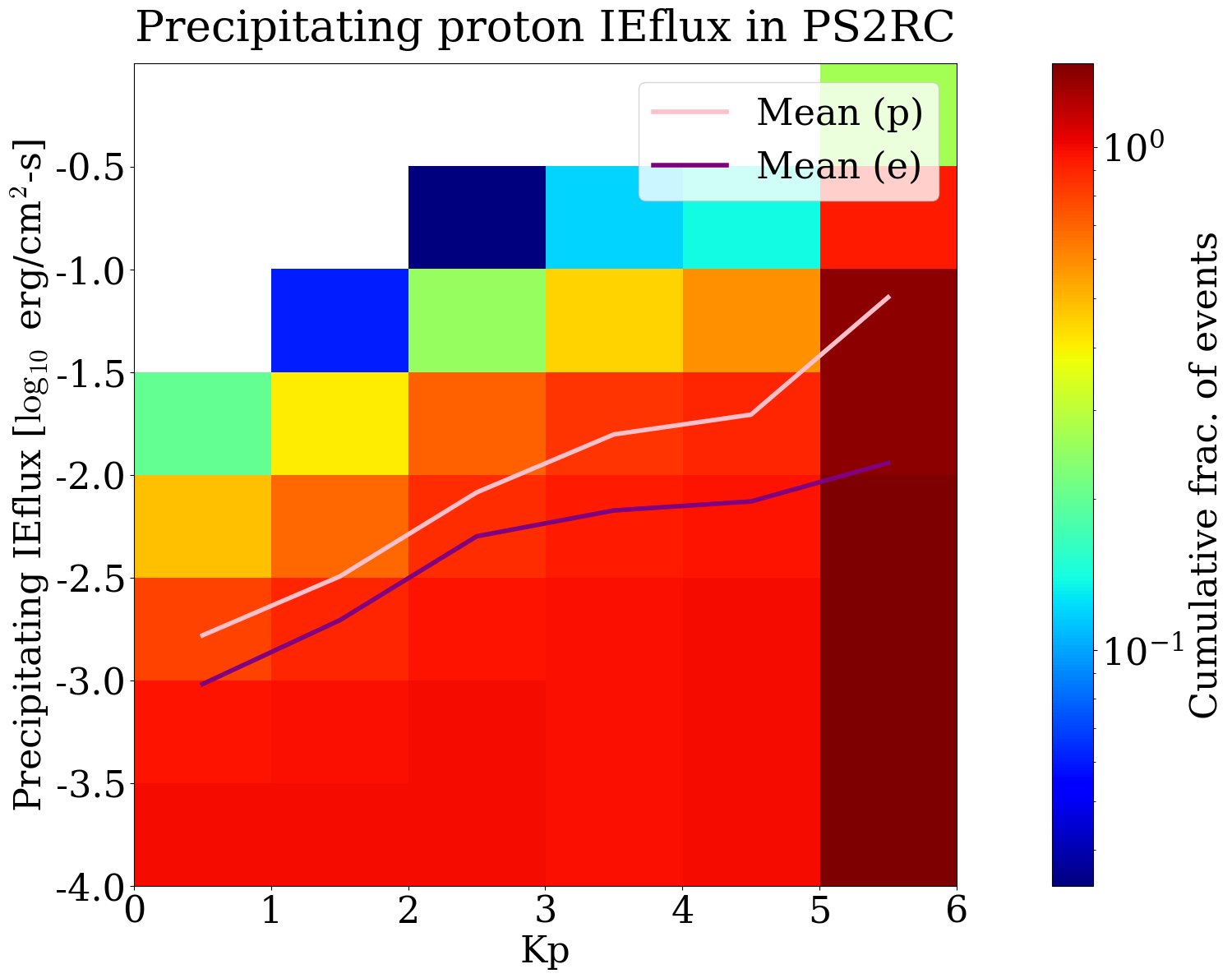}\\

\end{centering}
\caption{Precipitation ratio (left-column) and latitude-averaged integral precipitating energy-fluxes (IEfluxes; right-column) versus the geomagnetic activity indices AE (3 hour average; top), Dst (middle), and Kp (bottom). Quantities are equivalent to those described in Fig. 9. The precipitation ratio and average scale directly with activity indices, demonstrating that curvature scattering is a very important precipitation mechanism at active times on the nightside, including during storms and substorms. For protons, average precipitation can exceed 10$^0$ erg/cm$^2$-s in the PS2RC, with precipitation ratios approaching 100\% of high-latitude $\geq$50 keV contributions. Electrons typically contribute roughly 1/3 of the average proton precipitation, with the total storm-time precipitation being up to a factor of 10 lower. The average electron IB precipitation ratio is also under 50\% under all conditions, implying that mechanisms other than FLC can dominate the total electron high-latitude precipitation at quiet and active times.}
\label{Figure 10.}
\end{figure}


\end{document}